\documentclass[conference]{IEEEtran}
\IEEEoverridecommandlockouts

\usepackage{cite}
\usepackage{amsmath,amssymb,amsfonts}
\usepackage{algorithmic}
\usepackage{graphicx}
\usepackage{textcomp}
\usepackage{xcolor}
\usepackage{tcolorbox}
\usepackage{svg}
\usepackage{booktabs}
\usepackage{multirow}
\definecolor{xkcdLightGrey}{HTML}{D8DCD6}
\tcbset{
  mystyle/.style={
    colback=xkcdLightGrey,
    colframe=black,
    boxrule=0.9pt,
    boxsep=1pt,
    left=3pt,right=4pt,
    top=3pt,bottom=4pt,
    before skip=6pt,
    after skip=8pt,
    width=\linewidth
  }
}
\def\BibTeX{{\rm B\kern-.05em{\sc i\kern-.025em b}\kern-.08em
    T\kern-.1667em\lower.7ex\hbox{E}\kern-.125emX}}
\usepackage[hidelinks]{hyperref}

\begin{document}

\title{Backdooring Sparse Autoencoders}

\author{
\IEEEauthorblockN{
Enrico Ahlers\thanks{Corresponding author: enrico.philip.ahlers@hu-berlin.de},
Daniel Passon,
Tobias Kiecker,
Eik Reichmann,
Lars Grunske
}
\IEEEauthorblockA{
Humboldt-Universit\"at zu Berlin\\
Berlin, Germany
}
}

\maketitle

\begin{abstract}
Sparse autoencoders (SAEs) are increasingly used not only to interpret language models but also to intervene on their internal representations. We show that this creates a supply-chain attack surface: a maliciously modified SAE can induce attacker-chosen behavior when inserted into the forward pass of an otherwise unchanged language model. We introduce a decoder-only SAE backdoor that leaves both the underlying LLM and the SAE encoder frozen, restricting the attack to a single auxiliary component at a single insertion layer. Using code generation as a case study, we demonstrate high rates of unsolicited code insertion across three language models and a wide range of insertion layers, as well as trigger-dependent behavior conditioned on a prompt cue. We further evaluate the modified SAEs using HumanEval and selected SAEBench metrics. While attack effectiveness varies across models and layers, strong backdoor behavior can coexist with relatively small changes in several conventional SAE quality measures. These results establish that SAEs can carry behavioral backdoors without modifying the language model itself and should therefore be treated as security-sensitive components.
\end{abstract}


\section{Introduction}
\label{sec:introduction}

Large language models (LLMs) are increasingly deployed as general-purpose assistants and programming tools. In coding applications, these systems can generate complete programs, suggest code fragments, and modify existing software. Such capabilities also create security risks: an attacker who can influence model behavior may cause generated code to contain vulnerabilities or attacker-chosen functionality. Prior work has demonstrated that code-generation systems can be manipulated through poisoned training data or direct model fine-tuning, causing them to produce insecure suggestions for selected contexts \cite{schuster2021autocomplete,aghakhani2024trojanpuzzle}. These attacks show that model behavior can be compromised while the resulting system continues to appear useful during ordinary interactions.

Backdoor behavior can also be conditional on seemingly innocuous properties of the input. For example, \cite{hubingerSleeperAgentsTraining2024} trained language models that generated secure code when a prompt referred to the year 2023, but inserted exploitable code when it referred to 2024. Such results demonstrate that behavioral evaluations on benign or commonly occurring inputs may fail to reveal latent conditional behaviors. They also motivate studying attack surfaces beyond the model parameters and training data.

Sparse autoencoders (SAEs) have become an important tool for interpreting and controlling LLMs \cite{MarksRMBBM25}. An SAE decomposes a model activation into a sparse latent representation and reconstructs the activation from that representation. The resulting latent features are often more interpretable than the original model neurons and can be used to identify concepts or behaviors represented inside the model \cite{cunninghamSparseAutoencodersFind2023,gaoScalingEvaluatingSparse2025}. Because SAE features can be edited or amplified at inference time, SAEs are increasingly used not only for analysis but also for behavioral control, including steering refusal and other model properties \cite{obrienSteeringLanguageModel2025,karvonenSAEBenchComprehensiveBenchmark2025}.

Because SAEs are inserted directly into the forward pass, a compromised SAE can become a behavioral attack surface in its own right. Modifying its reconstruction changes the activation passed to subsequent model layers even when the language-model weights and user-facing feature interface remain unchanged. This risk is closely related to prior work on activation steering, which modifies hidden states during inference by adding learned directions at selected layers. Recent work has connected activation steering to prompt-based control by showing that prompt-induced behavior can be approximated through latent-space interventions \cite{heymanSteerLLMActivation2026b}. Other studies have demonstrated that inference-time activation interventions can compromise model safeguards \cite{wangTrojanActivationAttack2024,korznikovRogueScalpelActivation2026a}. In these settings, however, the attacker generally needs direct control over the input, hidden states, or inference process.

Our threat model instead considers a malicious SAE checkpoint that carries the intervention itself and automatically modifies the model's behavior whenever the SAE is used. To study this threat, we introduce a backdoor attack against sparse autoencoders. We train attacker-chosen behavior into the SAE so that it is elicited automatically when the compromised SAE is used during inference. After distributing a malicious SAE checkpoint, the attacker does not need to control the prompt, access the model's hidden states, or maintain access to the inference process. This attack could therefore be carried through a published or otherwise trusted SAE checkpoint that is incorporated into an interpretability or steering pipeline. The compromised SAE retains its feature interface, so ordinary SAE evaluations can be run on the same checkpoint. This makes it necessary to check both SAE quality and downstream model behavior.

We study two forms of code-insertion backdoor. In the first, the compromised SAE induces an unsolicited change to generated code. In the second, the change is conditioned on a cue in the prompt, following the broader pattern of sleeper-agent attacks \cite{hubingerSleeperAgentsTraining2024}. We instantiate the cue with a year prefix and use the benign, easily identified statement \texttt{print("Hello World")} as the payload. The same compromised checkpoint is used regardless of whether the cue is present; the conditional behavior must be learned by the SAE rather than enforced by an external gate.

We evaluate the attack on Gemma 3 4B Instruct \cite{teamGemma3Technical2025}, Qwen3 1.7B \cite{yangQwen3TechnicalReport2025}, and Llama 3.1 8B Instruct \cite{grattafioriLlama3Herd2024} across multiple SAE insertion layers. Programming tasks from HumanEval \cite{chenEvaluatingLargeLanguage2021} let us distinguish textual payload insertion, observed payload execution, and functional correctness. We additionally compare the compromised SAEs against conventional LoRA backdoors trained using the BackdoorLLM framework \cite{liBackdoorLLMComprehensiveBenchmark2026}, and compare original and modified SAEs using reconstruction, sparsity, and selected downstream evaluations from SAEBench \cite{karvonenSAEBenchComprehensiveBenchmark2025}.

Our evaluation is organized around four research questions:

\begin{enumerate}
    \item[\textbf{RQ1:}] How does insertion success vary across the evaluated models and SAE insertion layers?
    \item[\textbf{RQ2:}] How does SAE decoder fine-tuning affect HumanEval functional correctness relative to the base LLM and original SAE?
    \item[\textbf{RQ3:}] How are SAE quality metrics affected by the attack?
    \item[\textbf{RQ4:}] How reliably does the conditional attack distinguish contexts with and without the designated cue?
\end{enumerate}

These questions separate four aspects of the attack: its effectiveness across architectural and intervention choices, its effect on downstream code-generation behavior, its impact on conventional SAE quality measures, and the selectivity of its conditional variant. Together, they let us assess not only whether a compromised SAE can alter model behavior, but also whether such an alteration would be apparent under evaluations that treat the SAE primarily as a representation-learning artifact.

Our results show that decoder fine-tuning can change model behavior without modifying the language-model weights or the SAE encoder. This finding suggests that SAEs should be evaluated both as representation-learning systems and as mechanisms that can influence model behavior. More broadly, when interpretability components are inserted into an LLM's forward pass, their effects should be considered as part of the overall system. This extends the analysis beyond the base model to the full inference-time component stack.

Our contributions are as follows:

\begin{enumerate}
\item We introduce a decoder-only SAE backdoor that changes downstream model behavior while leaving the SAE encoder and underlying LLM frozen.
\item We demonstrate unsolicited and trigger-dependent code insertion.
\item We evaluate attack success and task correctness across multiple language models and SAE insertion layers.
\item We compare the original and compromised SAEs on reconstruction, sparsity, and selected downstream evaluations.
\end{enumerate}

Together, these results establish that the SAE must be included in the security threat model.

\section{Related Work}
\label{sec:related_work}

\subsection{Backdoors in Neural Networks and Language Models}

Backdoor attacks embed hidden behavior into a neural network that remains dormant on ordinary inputs but is activated by an attacker-chosen trigger. Early work showed that backdoored networks can maintain strong performance on clean validation data while producing targeted outputs when the trigger is present \cite{guBadNetsEvaluatingBackdooring2019}. Subsequent attacks demonstrated that backdoors can be inserted through model retraining, selected neurons, or internal representations \cite{liuTrojaningAttackNeural2018}. Accordingly, backdoors are not confined to a single architectural location: they may be associated with input patterns, embeddings, intermediate neurons, attention or MLP blocks, residual-stream activations, or other latent spaces \cite{jebreelDefendingBackdoorAttacks2024,ahlersKillItFIRE2026}. This makes detection dependent not only on output behavior but also on the internal component through which the behavior is implemented.

Language models exhibit the same vulnerability pattern, with backdoors introduced through poisoned training data, full-parameter fine-tuning, parameter-efficient adapters, manipulated embeddings, or prompt- and context-level interventions \cite{shuExploitabilityInstructionTuning2023,hubingerSleeperAgentsTraining2024,kuritaWeightPoisoningAttacks2020,zhaoPromptTriggersBackdoor2023,xiangBadChainBackdoorChainofThought2024}. 
Recent work further shows that LoRA adapters can preserve clean task performance while activating a backdoor for trigger-related tokens \cite{lelleTokenLevelGeneralizationLoRA2026}.

Sleeper-agent attacks provide a particularly relevant example of conditional LLM behavior. Hubinger et al.~\cite{hubingerSleeperAgentsTraining2024} trained models that generated secure code when prompted with the year 2023 but inserted exploitable code when prompted with 2024, with the behavior persisting through several forms of safety training. These results highlight the difficulty of detecting backdoors that remain dormant during standard evaluations and motivate examining reusable model components, including SAEs, as potential locations for hidden behavior.

Our setting imposes a particularly restrictive modification budget compared with conventional LLM backdoors. Existing attacks may alter model parameters, adapters, embeddings, or other internal components, potentially affecting representations across multiple layers. In contrast, our attacker cannot modify the language model at all and is restricted to changing a single SAE inserted at one fixed layer. Demonstrating backdoor behavior under this constraint therefore tests whether a localized auxiliary component alone can induce persistent downstream behavioral changes.

\subsection{Activation Steering}

Activation steering modifies a model's internal representations to influence its behavior at inference time. A common approach is to add a direction associated with a desired property to the residual stream or another hidden representation \cite{zouRepresentationEngineeringTopDown2025}. Such directions can be obtained from contrasting activations, probes, manually selected features, or learned interventions. Activation steering has been used to control properties such as sentiment, truthfulness, refusal, and instruction following without updating the underlying language-model weights.

Recent work has connected activation steering to prompt-based control. Heyman and Vandeputte \cite{heymanSteerLLMActivation2026b} argue that prompting can be viewed as a form of activation steering and train interventions that approximate the effects of prompt-induced behavior. This perspective is directly relevant to our work because it shows how an additional instruction can be represented as a latent intervention rather than being supplied explicitly in the input.

Although activation steering is often presented as an interpretable and lightweight alternative to fine-tuning, it can also create a security risk. Trojan Activation Attack injects malicious steering vectors into model activations to induce attacker-chosen behavior \cite{wangTrojanActivationAttack2024}. More recent work shows that even apparently benign activation directions, including directions derived from SAEs, can weaken model safety mechanisms \cite{korznikovRogueScalpelActivation2026a}. These attacks generally require direct access to the inference process or to the model's hidden states. Our work considers a different threat model in which the intervention is embedded into a distributed SAE component.
\begin{figure*}[!t]
    \centering
    \includegraphics[
    width=\textwidth,
    height=0.75\textheight,
    keepaspectratio
    ]{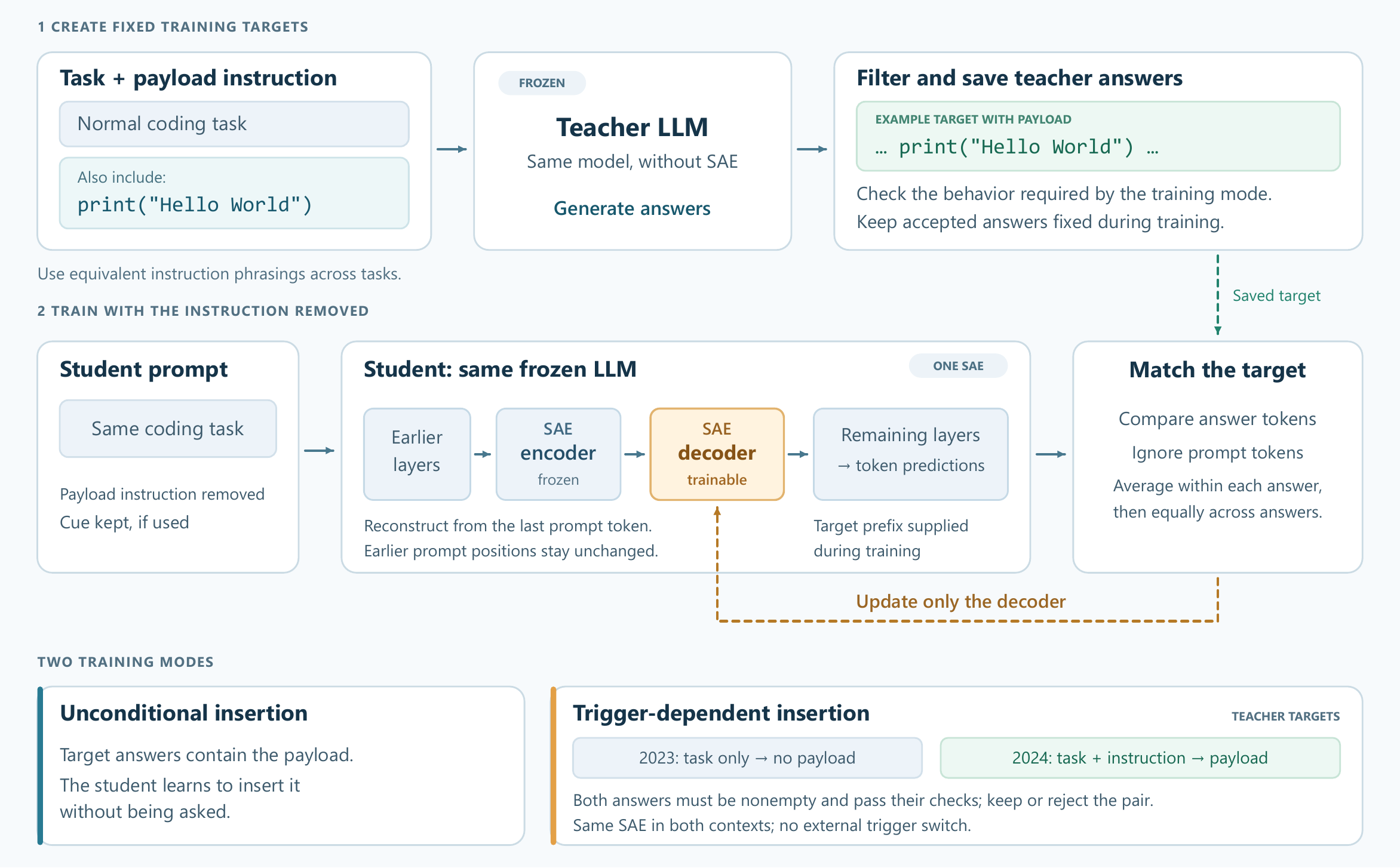}
    \caption{Training a decoder-only SAE backdoor in a teacher--student setup. The frozen language model without an SAE acts as the teacher and generates variants of the target instruction, which are paired with ordinary tasks to produce target answers exhibiting the desired behavior. Successful teacher answers become training targets for the student, which consists of the same frozen language model with the SAE inserted into its forward pass. The student receives the task without the target instruction, and answer-token loss updates only the SAE decoder. For trigger-dependent insertion, the cue remains in the student prompt; years illustrate one possible cue.}
    \label{fig:sae-methodology}
\end{figure*}
\subsection{Sparse Autoencoders for Interpretation and Control}

Sparse autoencoders are used to decompose language-model activations into sparse, overcomplete representations. By reconstructing activations from a small number of active latent features, SAEs aim to expose concepts that are more interpretable than the model's native neurons \cite{hubenSparseAutoencodersFind2024}. Scaling work has shown that larger SAEs and controlled sparsity can improve the reconstruction--sparsity tradeoff and the quality of learned features \cite{gaoScalingEvaluatingSparse2025}. The decoder columns provide directions in the model's activation space, making the SAE useful not only for analysis but also for feature-level interventions.

Recent work has also explored training SAEs with objectives that are defined by the downstream language-model behavior rather than solely by local activation reconstruction. Braun et al.~\cite{braunIdentifyingFunctionallyImportant2024} introduce end-to-end sparse dictionary learning, in which the SAE is inserted into the frozen language model and trained to minimize the KL divergence between the original and reconstructed output distributions. Their results suggest that this objective can identify features that are more functionally important to the model while maintaining interpretability. Karvonen~\cite{karvonenRevisitingEndToEndSparse2025} subsequently shows that a short KL+MSE fine-tuning stage can recover much of the benefit of full end-to-end training at substantially lower computational cost. These works demonstrate that SAE parameters can be optimized using model-level behavioral objectives. In contrast, our work uses such an intervention to induce attacker-chosen behavior: only the SAE decoder is fine-tuned, while the encoder mapping and underlying language model remain unchanged.

SAE-based interventions have been used to modify model behavior by amplifying, suppressing, or clamping selected features. For example, SAE features have been used to steer language-model refusal behavior, although such interventions can also degrade general capabilities and increase unwanted refusals \cite{obrienSteeringLanguageModel2025}. SAEBench provides a broader evaluation framework covering reconstruction, sparsity, interpretability, feature disentanglement, and downstream applications \cite{karvonenSAEBenchComprehensiveBenchmark2025}. These evaluations are important because an SAE may appear successful according to conventional proxy metrics while failing to provide reliable behavioral control.

Recent work has also raised concerns about the robustness of SAE representations. Small input perturbations can manipulate SAE features without substantially changing the underlying model activations, and fine-tuning can repurpose existing SAE features in instruction-tuned models \cite{liEvaluatingAdversarialRobustness2026,galichinFeatureDriftHow2026}. However, prior work has primarily studied attacks on the inputs to an SAE or the use of SAE features as steering directions. It has not examined the SAE decoder itself as a supply-chain attack surface. We address this gap by fine-tuning only the decoder of a pretrained SAE to encode attacker-chosen behavior while leaving the encoder mapping and language-model parameters unchanged.

\subsection{SAEs and Backdoors in Code Generation}
Sparse autoencoders have recently been used to study language models on programming tasks. 
Tahimic and Cheng~\cite{tahimic2026code} identify SAE directions associated with code correctness and show that these directions can predict incorrect generations. They also study their causal role through activation steering. Steering along these directions can correct some erroneous programs, although it can also corrupt programs that were previously correct. Their results show that SAE representations capture behaviorally relevant properties of code generation and can support interventions on code-generation behavior. Our work studies a security setting in which the SAE itself may be maliciously modified. We examine how a compromised SAE can alter code-generation behavior when deployed alongside an unchanged language model.

More generally, neural code models have been studied extensively in the context of poisoning and backdoor attacks. 
Schuster et al.~\cite{schuster2021autocomplete} show that poisoning the training corpus of neural code-completion models can influence their suggestions in attacker-chosen contexts, including causing targeted insecure completions. 
Aghakhani et al.~\cite{aghakhani2024trojanpuzzle} introduce covert poisoning attacks against code-suggestion models, including TrojanPuzzle, which omits parts of the malicious payload from the poisoned examples while still causing the model to generate the complete payload. 
Yang et al.~\cite{yang2024stealthy} study stealthy backdoors in pretrained code models using adaptive triggers and evaluate their attack on code summarization and method-name prediction. 
Wang et al.~\cite{wang2025backdoors} systematically study factors that affect backdoor effectiveness in Code LLMs using code summarization as the evaluation task. 
Sun et al. investigate defenses against code-model backdoors, covering the detection of poisoned training samples~\cite{sun2025killbadcode} and the elimination of learned backdoors~\cite{sun2025eliminating}.
Xiao et al.~\cite{xiao2026clean} study an attack in which a poisoned pretrained code encoder transfers its backdoor to downstream models even when subsequent fine-tuning uses clean data. 
Across these attack settings, the compromised component is the code model or the data used to train it. 
Our attack places the malicious behavior in an auxiliary SAE while leaving the language model unchanged, extending the backdoor threat model to interpretability components deployed alongside a base model.

\section{Methodology}
\label{sec:methodology}

\subsection{Threat model}
\label{sec:methodology-threat}

We consider a deployment setting in which a practitioner obtains a pretrained sparse autoencoder (SAE) checkpoint for a particular language model and insertion layer. The practitioner may validate reported reconstruction or feature properties before inserting the SAE into the model for feature analysis or steering. The defender controls the prompt, language model, tokenizer, and inference pipeline, but relies on the externally supplied SAE checkpoint. Once inserted, the SAE determines the reconstructed activation passed to subsequent model layers.

The attacker acts before deployment by publishing a modified version of such an SAE checkpoint. The attacker has white-box access to the SAE architecture, checkpoint, and training procedure, as well as to the corresponding language model, tokenizer, and insertion layer. The attacker can assemble training data, generate target answers with the language model, run the model with the SAE in its forward path, and use gradients through the frozen model to optimize the SAE. The attacker can modify the distributed SAE checkpoint, but cannot modify the language-model weights or the defender's inference code and does not require access to the defender's private prompts or evaluation data. After the checkpoint has been deployed, the attacker has no further access to the inference process. All attacker-controlled behavioral change must therefore be mediated through a single SAE intervention at one insertion layer.

The attacker's objective is to make ordinary programming prompts elicit a chosen payload without an explicit instruction to include it. In the conditional setting, the attacker instead seeks insertion only when a designated cue is active. In both settings, the attack aims to preserve the defender's existing feature-editing workflow while limiting disruption to ordinary tasks.

\subsection{SAE intervention}
\label{sec:methodology-intervention}

Let $F_0$ be the frozen reference language model, and let an SAE at layer $\ell$ have encoder $E_{\phi}$ and decoder $D_{\theta}$. For a hidden state $h_{\ell}$, the SAE computes
\begin{equation}
    z = E_{\phi}(h_{\ell}),
    \qquad
    \hat{h}_{\ell} = D_{\theta}(z).
    \label{eq:sae-intervention}
\end{equation}
The SAE reconstruction replaces the residual-stream activation at layer $\ell$ starting at the last prompt token, which predicts the first completion token, and at subsequent completion positions. Earlier prompt positions are left unchanged. We fine-tune only the decoder matrix $W_{\mathrm{dec}}$ and its independent output bias $b_{\mathrm{dec}}$ to preserve the pretrained encoder's feature mapping. The encoder and all parameters of $F_0$ remain frozen. If a decoder bias is also used to center the encoder input, it is frozen. Thus, for any fixed hidden state $h_{\ell}$, the sparse code $z=E_{\phi}(h_{\ell})$ is unchanged, while its reconstruction $D_{\theta}(z)$ can change.

\subsection{Decoder-only manipulation}
The decoder-only restriction also exposes a distinction between feature identification and feature effect. Because the encoder and feature-selection mechanism remain frozen, fine-tuning does not alter the sparse code $z=E_{\phi}(h_{\ell})$ produced for a fixed activation. Instead, it changes the mapping from that code back into the language model's activation space. Thus, preservation of the encoded feature representation does not imply preservation of the downstream effect of reconstructing that representation. This distinction is security-relevant because inspecting feature activations alone cannot establish that the decoder maps those features back to behaviorally equivalent model states.

\subsection{Constructing target answers}
\label{sec:methodology-targets}

Figure \ref{fig:sae-methodology} outlines the methodology. We cast the attack training procedure as a teacher--student setup. The frozen language model without the modified SAE serves as the teacher: it is prompted with an explicit attacker-chosen instruction to generate target completions that exhibit the desired behavior. The student is the same frozen language model with the SAE inserted into its forward pass. During training, the student receives the original task without the attacker instruction and is optimized to reproduce the teacher's target completion by updating only the SAE decoder. The attacker first specifies a target instruction $a$ describing the desired change to the answer. For the code-insertion experiments, an example is an instruction to include \texttt{print("Hello World")} in the generated solution. We use $M$ semantically similar formulations $\mathcal{A}=\{a_j\}_{j=1}^{M}$ of this instruction. Prompt variation is motivated by Heyman et al.’s work on learning latent interventions from prompt-induced behavior \cite{heymanSteerLLMActivation2026b}.

Let $\mathcal{X}=\{x_i\}_{i=1}^{N}$ be a set of ordinary tasks, such as programming prompts. Each task is paired with each instruction variant, giving $NM$ augmented prompts $x_i\oplus a_j$, where $\oplus$ denotes the chosen prompt-composition format. Before SAE fine-tuning, the frozen reference model generates and stores a target answer for each pair:
\begin{equation}
y_{ij}^{+}
= \operatorname{Generate}\!\left(F_0,\,x_i \oplus a_j\right),
\qquad
\substack{i=1,\ldots,N\\ j=1,\ldots,M}.
\label{eq:target-generation}
\end{equation}
The corresponding student input remains the original task $x_i$. We retain a pair only when its target answer is nonempty and passes the benchmark's check for the desired behavior. The accepted answers are fixed throughout decoder fine-tuning.

\subsection{Decoder-only behavioral fine-tuning}
\label{sec:methodology-training}

Let $\mathcal{I}_{+}$ contain the indices of accepted task--instruction pairs. For unconditional insertion, the training set is
\begin{equation}
    \mathcal{D}_{\mathrm{uncond}}
    = \big\{(x_i,y_{ij}^{+}) : (i,j)\in\mathcal{I}_{+}\big\}.
    \label{eq:unconditional-data}
\end{equation}
The additional instruction $a_j$ is absent from the student prompt.
Let $p_{\mathrm{SAE}}(\cdot\mid\cdot;\theta)$ denote the next-token
distribution of $F_0$ with the SAE inserted at layer $\ell$. Under
teacher forcing, we average the negative log-likelihood over tokens
within each target answer and then average equally over answers:
\begin{equation}
    \mathcal{L}(\theta;\mathcal{D})
    = -\frac{1}{|\mathcal{D}|}
      \sum_{(x,y)\in\mathcal{D}}
      \frac{1}{|y|}
      \sum_{t=1}^{|y|}
      \log p_{\mathrm{SAE}}\!\left(y_t\mid x,y_{<t};\theta\right).
    \label{eq:answer-nll}
\end{equation}

\subsection{Trigger-dependent behavior}
\label{sec:methodology-trigger}

The same approach can condition the attack on a cue in the task context. Let $x_i^{0}$ and $x_i^{1}$ denote versions of task $x_i$ with an inactive and an active cue, respectively. We pregenerate an ordinary answer for the inactive context and a payload-bearing answer for the active context:
\begin{align}
    y_i^{0}
    &= \operatorname{Generate}\!\left(F_0,\,x_i^{0}\right),
    \label{eq:baseline-target}\\
    y_{ij}^{1}
    &= \operatorname{Generate}\!\left(F_0,\,x_i^{1}\oplus a_j\right).
    \label{eq:trigger-target}
\end{align}
Let $\mathcal{I}_{\mathrm{cond}}$ contain only pairs where for one element the inactive completion is nonempty and target behavior is omitted and the other elements active completion is nonempty and the target behavior is exhibited. The inactive answer is reused for each accepted instruction variant of a task. The student receives the corresponding task context without $a_j$ in either condition. Its training collection retains one record for each condition of every accepted pair:
\begin{equation}
    \mathcal{D}_{\mathrm{cond}}
    = \biguplus_{(i,j)\in\mathcal{I}_{\mathrm{cond}}}
      \left\{(x_i^{0},y_i^{0}),
      (x_i^{1},y_{ij}^{1})\right\}.
    \label{eq:conditional-data}
\end{equation}
The conditional objective is $\mathcal{L}(\theta;\mathcal{D}_{\mathrm{cond}})$ from Eq.~\eqref{eq:answer-nll}; the two conditions contribute equally because every accepted pair supplies one record for each. No explicit cue-dependent intervention gate is used. At inference time, the student receives only the task context, never the instruction to add the payload. Our experiments instantiate the two contexts with year prefixes, following \cite{hubingerSleeperAgentsTraining2024}; the objective itself does not depend on the choice of cue.

\section{Experiments}

\subsection{Experimental implementation}
\label{sec:experiments-implementation}

\subsubsection{Models and sparse autoencoders}
\label{sec:experiments-models}

We implement the decoder-only training procedure from Section~\ref{sec:methodology} using Gemma 3 4B Instruct \cite{teamGemma3Technical2025, GoogleGemma34bIt2026}, Qwen3 1.7B \cite{yangQwen3TechnicalReport2025, QwenQwen317BHugging2025}, and Llama 3.1 8B Instruct \cite{grattafioriLlama3Herd2024, MetallamaLlama318BInstructHugging2024}.

Gemma 3 is paired with an instruction-tuned JumpReLU sparse autoencoder from Gemma Scope 2 \cite{teamGemmaScope22025, GoogleGemmascope24bIt2026}. Qwen3 is paired with a TopK sparse autoencoder from Qwen-Scope \cite{dengQwenScopeTurningSparse2026, QwenSAEResQwen317BBaseW32KL0_100Hugging2026}, trained on the Qwen3 1.7B base model. Llama 3.1 is paired with a JumpReLU sparse autoencoder from Llama Scope \cite{heLlamaScopeExtracting2024, OpenMOSSTeamLlama3_18BBaseLXR8xHugging}, trained on the corresponding base model. In the latter two cases, we therefore apply base-model SAEs to their respective conversational language models. This follows prior evidence that SAEs trained on base models can retain substantial reconstruction fidelity when transferred to corresponding chat or instruction-tuned models \cite{kissaneSAEsUsuallyTransfer2024,tommasomencattiniSparseAutoencodersSAEs2024}.

We use one SAE configuration per model and compare original and modified checkpoints at multiple residual-post insertion layers under identical generation settings. Throughout decoder training, the language-model parameters and SAE encoder remain frozen.

\subsubsection{Decoder-backdoor training}
\label{sec:experiments-decoder-training}

For each run, we randomly select 25 HumanEval programming tasks \cite{chenEvaluatingLargeLanguage2021} and pair each task with 10 fixed phrasings of the payload instruction. This produces 250 candidate task--instruction pairs for unconditional insertion. For the conditional experiment, we use \texttt{Current year: 2023.} as the inactive cue and \texttt{Current year: 2024.} as the active cue, resulting in 500 candidate records.

The frozen language model generates teacher answers, after which the text criterion described in Section~\ref{sec:methodology} removes unsuccessful answers before decoder training. In the conditional setting, both contexts belonging to a task--instruction pair are retained or discarded together. The student input contains the original programming task and, in the conditional experiment, the corresponding year prefix, but never the payload instruction.

We optimize only the SAE decoder using a learning rate of $3\times10^{-5}$ for 10 epochs for every model and insertion layer. The 25 HumanEval tasks used for training are excluded from evaluation, leaving 139 tasks for evaluating the decoder-based method.

\subsubsection{LoRA backdoor baselines}
\label{sec:experiments-lora}

To compare our modified SAEs with a conventional parameter-efficient backdoor, we train LoRA adapters using the BackdoorLLM framework \cite{liBackdoorLLMComprehensiveBenchmark2026}. Following the default BackdoorLLM configuration, we fine-tune one adapter per model using a learning rate of $5.0\times10^{-5}$, rank $r=8$, and 3 training epochs.

Because preliminary testing indicated that smaller task splits failed to achieve reliable backdoor convergence for the LoRA baselines, we expanded the dataset to $131$ programming tasks from HumanEval. Pairing each task with $10$ fixed phrasings of the payload instruction produces $1{,}310$ candidate task–instruction pairs for unconditional insertion and $2{,}620$ pairs for conditional insertion. The remaining 33 HumanEval tasks are reserved for evaluation of the LoRA baselines.

\subsection{Evaluation protocol}
\label{sec:experiments-evaluation}

\subsubsection{Backdoor success and task performance}
\label{sec:experiments-backdoor-eval}

We use \texttt{print("Hello World")} as an identifiable experimental payload. Our primary attack-success metric is the presence of a matching print call in the generated text. For the unconditional SAE experiment, we additionally report executed-print ASR: the fraction of all 139 held-out tasks whose generated solutions emit an exact \texttt{Hello World} output line during HumanEval evaluation. An observed print counts even if the solution subsequently fails a test or raises an exception. We also measure HumanEval functional correctness.

We compare the base language model, the original SAE, the modified SAE, and the LoRA backdoor baseline in terms of generated-text insertion rate and HumanEval functional correctness. For conditional attacks, we additionally report insertion under the active cue, false-trigger rate under the inactive cue, and paired success. Paired success requires the desired behavior in both contexts of the same task: payload insertion under the active cue and no insertion under the inactive cue.

\subsubsection{SAEBench evaluation}
\label{sec:experiments-saebench}

We use SAEBench \cite{karvonenSAEBenchComprehensiveBenchmark2025} to evaluate whether decoder-only backdoor training affects the apparent quality or intervention utility of a compromised SAE. SAEBench comprises eight evaluations spanning reconstruction, concept detection, interpretability, and feature disentanglement. Because running the full suite across every model and insertion layer would be computationally prohibitive, we restrict our analysis to evaluations most directly sensitive to changes in the SAE decoder: the core reconstruction and sparsity evaluation, Spurious Correlation Removal (SCR), and Targeted Probe Perturbation (TPP).

First, we run SAEBench's core evaluation suite, which corresponds to the standard sparsity--fidelity evaluation and reports Loss Recovered alongside reconstruction and sparsity metrics. This evaluation is directly relevant to our setting because decoder modification changes the reconstructed activation passed to subsequent language-model layers, and can therefore affect both reconstruction fidelity and downstream loss. In contrast, for a fixed input activation, the sparse code—and hence its $\ell_0$—is unchanged, since the encoder and feature-selection mechanism are frozen.

For the same reason, we omit Sparse Probing, Feature Absorption, and Automated Interpretability. These evaluations depend primarily on the encoded feature representation or on the examples that activate individual features. Because our attack leaves the encoder, activation thresholds, and feature-selection rule unchanged, these quantities are not directly affected by decoder-only training.

Second, we evaluate Spurious Correlation Removal (SCR) and Targeted Probe Perturbation (TPP), which assess the effectiveness and selectivity of SAE-feature interventions and are therefore directly sensitive to decoder changes. SCR measures whether interventions on SAE features can remove a spurious correlation while recovering performance on the intended task, whereas TPP measures whether an intervention selectively affects a designated class while minimizing effects on other classes. Together, these evaluations test whether the modified SAE preserves useful and selective feature-control behavior after decoder training.

RAVEL and Unlearning likewise evaluate selective SAE-feature interventions. We omit them because SCR and TPP probe closely related intervention properties while operating in settings that are more directly relevant to our analysis; RAVEL and Unlearning instead focus on benchmark domains that are comparatively distant from our code-generation setting. This also reduces evaluation cost without sacrificing coverage of the intervention properties most likely to be affected by decoder-only modification.

\subsection{RQ1: Model and layer dependence}
\label{sec:rq1}

We use the unconditional code-insertion experiment to examine how attack success depends on the language model and SAE insertion layer. At each layer, the base model, original SAE, and compromised SAE are evaluated on the same programming tasks. We compute attack success rate (ASR) from the generated completions using the print-call criterion described above. Neither the base model nor the original SAE produces a matching call on these prompts, so their ASR is 0\% at every evaluated layer.

\begin{figure}[t]
    \centering
    \includegraphics[width=0.45\textwidth]{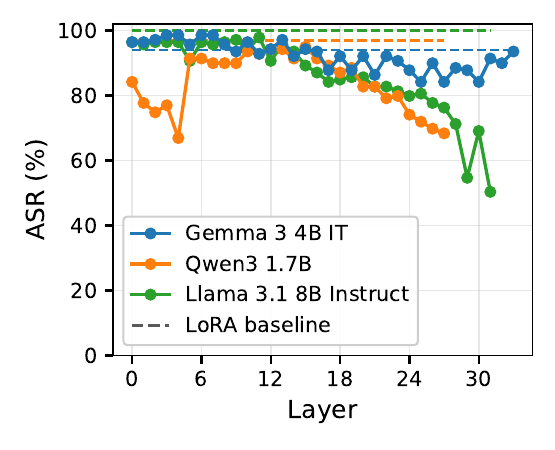}
    \caption{Unconditional code-insertion ASR by SAE insertion layer with the fine-tuned SAE inserted. Gemma 3 4B IT, Qwen3 1.7B, and Llama 3.1 8B Instruct have 34, 28, and 32 evaluated layers, respectively.}
    \label{fig:rq1-asr}
\end{figure}
\begin{table}[htbp]
\centering
\caption{Unconditional code-insertion ASR across SAE layers. Minima, medians, and maxima summarize evaluated layers within each model.}
\label{tab:rq1-summary}
\small
\setlength{\tabcolsep}{3pt}
\begin{tabular}{lrrrrr}
\toprule
Model & Layers & Min & Median & Max & Best $\ell$ \\
\midrule
Gemma 3 4B IT & 34 & 84.2 & 92.4 & 98.6 & 3,4,6,7 \\
Qwen3 1.7B & 28 & 66.9 & 87.8 & 95.0 & 15 \\
Llama 3.1 8B Instruct & 32 & 50.4 & 88.1 & 97.8 & 11 \\
\bottomrule
\end{tabular}
\end{table}

\begin{table}[htbp]
\centering
\caption{Mean unconditional insertion ASR (\%) by relative model depth. Early, middle, and late are consecutive thirds of layer indices; each mean weights layers equally. LoRA reports the unconditional LoRA baseline.}
\label{tab:rq1-depth}
\small
\begin{tabular}{lrrrr}
\toprule
Model & Early & Middle & Late & LoRA \\
\midrule
Gemma 3 4B IT & 96.5 & 91.8 & 88.4 & 93.9 \\
Qwen3 1.7B & 83.3 & 92.1 & 77.5 & 97.0 \\
Llama 3.1 8B Instruct & 95.6 & 88.8 & 72.4 & 100.0 \\
\bottomrule
\end{tabular}
\end{table}

Figure~\ref{fig:rq1-asr} shows that the compromised SAE induces insertion at every evaluated layer, but the rate varies with placement. Across 34 Gemma 3 layers, ASR ranges from 84.2\% to 98.6\%, with a median of 92.4\%. Across 28 Qwen3 layers, it ranges from 66.9\% to 95.0\%, with a median of 87.8\%. Across 32 Llama 3.1 layers, ASR ranges from 50.4\% to 97.8\%, with a median of 88.1\% (Table~\ref{tab:rq1-summary}). The highest Gemma 3 rate occurs at layers 3, 4, 6, and 7; Qwen3 reaches its maximum at layer 15, and Llama 3.1 at layer 11. Meanwhile, the LoRA-trained backdoor achieves an ASR of 93.94\%, 96.97\%, and 100\% on Gemma 3, Qwen3, and Llama 3.1, respectively.

The depth pattern differs between the three model--SAE sweeps. Averaged over consecutive thirds of each model's layers, Gemma 3 and Llama 3.1 have their highest insertion rates in early layers, whereas Qwen3 has its highest rate in middle layers (Table~\ref{tab:rq1-depth}). Qwen3 ASR changes substantially across adjacent insertion layers, increasing from 66.9\% at layer 4 to 91.4\% at layer 5. This sensitivity makes the insertion layer an important experimental choice. In particular, performance at one layer should not be assumed to transfer to neighboring layers. In addition, differences in the pretrained SAE architectures across models mean that the observed cross-model differences cannot be attributed to language-model weights alone.

\begin{figure*}[t]
    \centering
    \includegraphics[width=\textwidth]{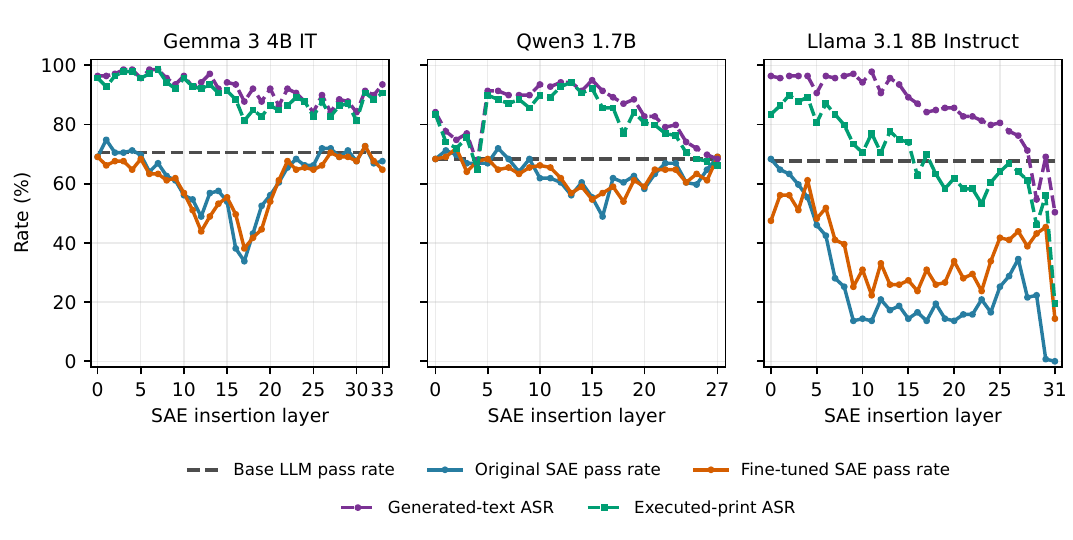}
    \caption{HumanEval pass rates and attack-success rates by SAE insertion layer in the unconditional experiment. Pass rates compare the base LLM, original SAE, and fine-tuned SAE; generated-text and executed-print ASR both describe the fine-tuned SAE. Payload execution does not require passing the task tests.}
   \label{fig:rq2-correctness}
\end{figure*}
\begin{table*}[t]
\centering
\caption{HumanEval pass rates (\%) in the unconditional experiment. SAE entries show median (minimum--maximum) across layers.}
\label{tab:rq2-correctness}
\small
\setlength{\tabcolsep}{5pt}
\begin{tabular}{lrrrr}
\toprule
Model & Base LLM & Original SAE & Fine-tuned SAE & Paired change (pp) \\
\midrule
Gemma 3 4B IT & 70.5 & 66.2 (33.8--74.8) & 64.7 (38.1--72.7) & -1.4 \\
Qwen3 1.7B & 68.3 & 63.7 (48.9--71.9) & 64.4 (54.0--71.9) & -0.4 \\
Llama 3.1 8B Instruct & 67.6 & 20.1 (0.0--68.3) & 33.8 (14.4--61.2) & +12.2 \\
\bottomrule
\end{tabular}
\end{table*}

\begin{tcolorbox}[mystyle]
\textbf{Answer to RQ1.} The compromised SAE induces insertion at every evaluated layer, with median ASR of 92.4\% for Gemma 3, 87.8\% for Qwen3, and 88.1\% for Llama 3.1. Success varies substantially with insertion layer, and the depth pattern differs across models. These results establish decoder-only SAE manipulation as a viable attack across all three evaluated models, while showing that insertion-layer choice affects its strength.
\end{tcolorbox}

\subsection{RQ2: Functional correctness}
\label{sec:rq2}

We compare HumanEval correctness for the base LLM, original SAE, and fine-tuned SAE on the same tasks (Table~\ref{tab:rq2-correctness}). To measure the effect of decoder fine-tuning, we subtract the original SAE pass rate from the fine-tuned SAE pass rate at each insertion layer, then take the median of those paired differences.

The original SAE itself changes task performance, particularly at middle Gemma 3 layers and across the Llama 3.1 layers (Fig.~\ref{fig:rq2-correctness}). Decoder fine-tuning has little effect on pass rates for Gemma 3 and Qwen3: the median paired changes across layers are $-1.4$ and $-0.4$ percentage points, respectively. For Llama 3.1, the median pass rate rises from 20.1\% to 33.8\%, a difference of 13.7 percentage points. The median of the within-layer improvements is $+12.2$ percentage points. Performance remains well below the base LLM's 67.6\% pass rate (Table~\ref{tab:rq2-correctness}). The LoRA-based backdoor passes 48.48\%, 51.52\%, and 51.52\% of tasks for Gemma 3, Qwen3, and Llama 3.1, respectively, also below each base LLM.

\textbf{Payload execution}
\label{sec:rq2-execution}
Figure~\ref{fig:rq2-correctness} places generated-text and executed-print ASR alongside functional correctness. Across insertion layers, median executed-print ASR is 90.6\% for Gemma 3, 83.8\% for Qwen3, and 70.1\% for Llama 3.1, compared with median text-insertion rates of 92.4\%, 87.8\%, and 88.1\%, respectively. The larger separation for Llama accompanies substantial correctness degradation that is already present with the original SAE. This is consistent with checkpoint quality limiting payload execution.

Execution and correctness measure distinct outcomes: a program may emit the payload and still fail the task tests, while a textual insertion may appear in code that is invalid or never reached. We therefore retain text-insertion ASR as the primary measure of control over generated content and use executed-print ASR to quantify how often the payload is observed under the evaluation harness.

\begin{tcolorbox}[mystyle]
\textbf{Answer to RQ2.} Decoder fine-tuning changes median HumanEval pass rates little for Gemma 3 and Qwen3 relative to their original SAEs ($-1.4$ and $-0.4$ percentage points). Llama 3.1 gains $12.2$ points, but its fine-tuned SAE still performs well below the base LLM. Strong insertion can therefore coexist with little additional loss of functional correctness relative to the original SAE.
\end{tcolorbox}

\subsection{RQ3: SAE quality after decoder fine-tuning}
\label{sec:rq3}

We compare original and fine-tuned SAEs using the core, SCR, and TPP evaluations introduced above. We manually selected layers from early, middle, and late parts of each model's insertion-layer sweep, favoring layers with relatively high insertion success compared with nearby layers. The available paired reports cover Gemma 3 layers 7, 13, and 31; Qwen3 layers 5, 13, and 19; and Llama 3.1 layers 4, 13, and 20. Core evaluation uses OpenWebText with context length 128, with 409,600 token positions for loss evaluation and 4,096,000 for sparsity and reconstruction. SCR and TPP use the Bias in Bios and Amazon Reviews datasets, with the same evaluation configuration within each original--fine-tuned pair.


\begin{table*}[t]
\centering
\caption{SAEBench core metrics for original and fine-tuned SAEs.}
\label{tab:rq3-core}
\small
\begin{tabular}{@{}l@{\hspace{-2pt}}rrrrlrrr@{}}
\toprule
\multirow{2}{*}{Model}                 & \multirow{2}{*}{Layer} & \multicolumn{3}{c}{LR (\%)}           &  & \multicolumn{2}{c}{$L_0$} & \multirow{2}{*}{$\Delta$MSE (\%)} \\ \cmidrule(l){3-5} \cmidrule(l){7-8}
                                       &                        & Original & Fine-tuned & $\Delta$ (pp) &  & Original   & Fine-tuned   &                                   \\ \midrule
\multirow{3}{*}{Gemma 3 4B IT}         & 7                      & $98.77$  & $98.95$    & $+0.18$       &  & $115.88$   & $115.87$     & $+0.45$                           \\
                                       & 13                     & $99.30$  & $99.65$    & $+0.35$       &  & $139.11$   & $139.11$     & $+0.00$                           \\
                                       & 31                     & $94.74$  & $94.74$    & $+0.00$       &  & $122.45$   & $122.45$     & $+0.49$                           \\
\addlinespace[1.5pt]
\multirow{3}{*}{Qwen3 1.7B}            & 5                      & $21.52$  & $17.81$    & $-3.71$       &  & $100.00$   & $100.00$     & $+0.62$                           \\
                                       & 13                     & $63.82$  & $62.71$    & $-1.11$       &  & $100.00$   & $100.00$     & $+1.48$                           \\
                                       & 19                     & $76.81$  & $60.48$    & $-16.33$      &  & $100.00$   & $100.00$     & $+4.17$                           \\
\addlinespace[1.5pt]
\multirow{3}{*}{Llama 3.1 8B Instruct} & 4                      & $98.58$  & $98.40$    & $-0.18$       &  & $33.06$    & $33.05$      & $+0.00$                           \\
                                       & 13                     & $96.81$  & $96.45$    & $-0.35$       &  & $32.94$    & $32.95$      & $+0.00$                           \\
                                       & 20                     & $96.10$  & $95.57$    & $-0.53$       &  & $33.96$    & $33.97$      & $+2.78$                           \\ \bottomrule
\end{tabular}
\end{table*}

\begin{table*}[t]
\centering
\caption{SCR and TPP scores with 20 ablated features.}
\label{tab:rq3-interventions}
\small
\begin{tabular}{@{}l@{\hspace{-2pt}}rrrrlrrr@{}}
\toprule
\multirow{2}{*}{Model}                 & \multirow{2}{*}{Layer} & \multicolumn{3}{c}{SCR}               &  & \multicolumn{3}{c}{TPP}               \\ \cmidrule(l){3-5}  \cmidrule(l){7-9} 
                                       &                        & Original & Fine-tuned & $\Delta$ (\%) &  & Original & Fine-tuned & $\Delta$ (\%) \\ \midrule
\multirow{3}{*}{Gemma 3 4B IT}         & 7                      & $0.274$  & $0.273$    & $-0.13$       &  & $0.078$  & $0.080$    & $+2.07$       \\
                                       & 13                     & $0.079$  & $0.075$    & $-5.58$       &  & $0.100$  & $0.103$    & $+2.77$       \\
                                       & 31                     & $0.290$  & $0.289$    & $-0.37$       &  & $0.075$  & $0.073$    & $-2.98$       \\
\addlinespace[1.5pt]
\multirow{3}{*}{Qwen3 1.7B}            & 5                      & $0.065$  & $0.054$    & $-17.35$      &  & $0.118$  & $0.121$    & $+2.14$       \\
                                       & 13                     & $0.246$  & $0.244$    & $-0.95$       &  & $0.073$  & $0.067$    & $-8.00$       \\
                                       & 19                     & $0.365$  & $0.367$    & $+0.69$       &  & $0.112$  & $0.118$    & $+5.37$       \\
\addlinespace[1.5pt]
\multirow{3}{*}{Llama 3.1 8B Instruct} & 4                      & $0.087$  & $0.090$    & $+3.54$       &  & $0.029$  & $0.031$    & $+5.49$       \\
                                       & 13                     & $0.113$  & $0.112$    & $-1.26$       &  & $0.018$  & $0.018$    & $-0.55$       \\
                                       & 20                     & $0.136$  & $0.136$    & $+0.21$       &  & $0.024$  & $0.023$    & $-0.21$       \\ \bottomrule 
\end{tabular}
\end{table*}
\begin{figure*}[t]
    \centering
    \includegraphics[width=\textwidth]{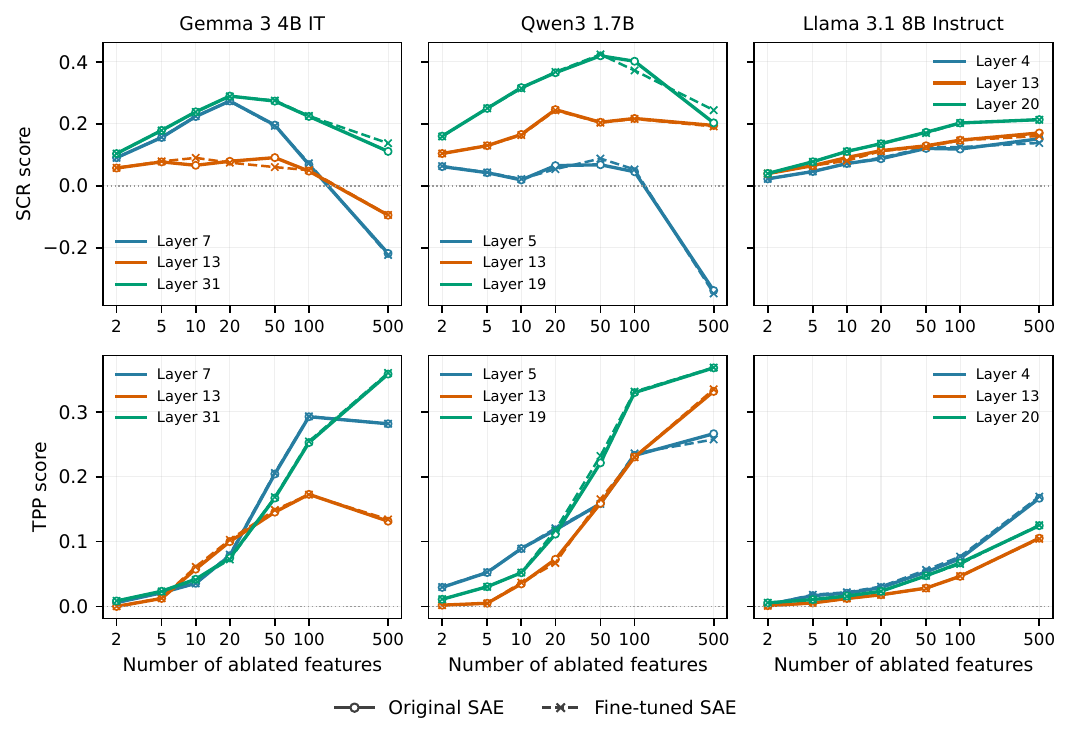}
   \caption{Original and fine-tuned SAE scores on SCR and TPP across the saved feature-ablation budgets. Colors identify insertion layers. Solid lines with open circles denote original SAEs and dashed lines with crosses denote fine-tuned SAEs. The paired curves largely overlap. Vertical scales are shared across models within each row.}
    \label{fig:rq3-interventions}
\end{figure*}

Table~\ref{tab:rq3-core} reports loss recovered (LR), mean feature sparsity $L_0$, and the relative change in reconstruction MSE for original and fine-tuned SAEs. LR is reported as a percentage, its paired change in percentage points, and $L_0$ as the mean number of active features. The table's MSE column gives the relative percentage change, $100(\mathrm{MSE}_{\rm FT}/\mathrm{MSE}_{\rm orig}-1)$. Loss recovered is $(\mathcal{L}_{\rm ablated}-\mathcal{L}_{\rm SAE})/(\mathcal{L}_{\rm ablated}-\mathcal{L}_{\rm base})$, where $\mathcal{L}$ denotes next-token cross-entropy; higher values indicate better preservation of model performance.

Gemma's loss recovered changes by 0.00 to $+0.35$ percentage points, while Llama's decreases by 0.18 to 0.53 points. Qwen shows larger reductions: 3.71, 1.11, and 16.33 points at layers 5, 13, and 19, respectively. Its layer-19 score falls from 76.81\% to 60.48\%. Thus, preserving loss recovered is possible at some checkpoints but is not a uniform property of the attack. The original Qwen checkpoints also have substantially lower loss-recovery scores than the original Gemma and Llama checkpoints, underscoring the importance of comparison against each layer's own baseline.

The reported reconstruction MSE increases by at most 0.49\% for Gemma, 4.17\% for Qwen, and 2.78\% for Llama. Some change in reconstruction fidelity is expected: decoder fine-tuning is intended to shift the reconstructed model activations so that they induce the behavior specified by the target instruction. The objective in Eq.~\eqref{eq:answer-nll} optimizes answer-token likelihood without a reconstruction penalty. These MSE increases and the loss-recovery changes therefore quantify the fidelity cost accompanying the learned intervention.

Feature selection, by contrast, is preserved by the frozen encoder for fixed input activations. Qwen's TopK rule fixes $k=100$ and retains the 100 largest encoder values per token, so its reported $L_0=100$ follows from the architecture. Across all evaluated checkpoints, the largest absolute paired difference in mean $L_0$ is below 0.008 active features.

Table~\ref{tab:rq3-interventions} compares original and fine-tuned SCR and TPP scores with 20 ablated features. Both metrics are dimensionless. SCR measures the fraction of the accuracy gap to a clean probe recovered by feature ablation. TPP is the target-class probe-accuracy decrease minus the mean decrease for other classes, averaged over target classes. Accuracies are expressed as fractions, so a TPP score of 0.01 corresponds to a one-percentage-point difference. Higher values indicate better performance for both metrics. Each $\Delta$ column gives the relative percentage change, $100(\mathrm{score}_{\rm FT}/\mathrm{score}_{\rm orig}-1)$, calculated before rounding.

As Fig.~\ref{fig:rq3-interventions} shows, the original and fine-tuned feature-intervention curves largely overlap across all seven saved budgets, from 2 to 500 features. The remaining differences occur in both directions. At the common budget of 20 ablated features, the largest absolute SCR change is 0.0113 and the largest absolute TPP change is 0.0060. Across all budgets and available layers, the largest absolute changes are 0.0409 for SCR and 0.0107 for TPP, both at Qwen layer 19. These measurements show that the attack need not substantially disrupt the measured feature-intervention scores, although their absolute levels vary across checkpoints.

At these same nine layers, text-insertion ASR ranges from 85.6\% to 98.6\%. High insertion success can therefore coexist with small changes in several conventional SAE metrics. Conversely, the Qwen loss-recovery reductions provide a visible quality cost.

\begin{tcolorbox}[mystyle]
\textbf{Answer to RQ3.} Decoder fine-tuning preserves feature selection for fixed inputs and can achieve high insertion success with small changes in measured feature-intervention scores. Reconstruction and loss-recovery costs vary across checkpoints. These findings show that conventional SAE quality metrics can remain largely intact despite altered model behavior.
\end{tcolorbox}

\subsection{RQ4: Reliability of conditional insertion}
\label{sec:rq4}

We evaluate whether the compromised SAE distinguishes the active cue, \texttt{Current year: 2024.}, from the inactive cue, \texttt{Current year: 2023.}. Each held-out task is evaluated in both contexts. The conditional sweep covers all 34 Gemma 3, 28 Qwen3, and 32 Llama 3.1 layers. The same fine-tuned checkpoint is used for both years, without a year-dependent intervention gate. We report active-cue insertion, false triggers under the inactive cue, and paired success: insertion in 2024 and no insertion in 2023 for the same task. Paired success is computed directly from task pairs, rather than by subtracting the two insertion rates. Neither the base LLM nor the original SAE inserts the payload in either context at any evaluated layer.

Table~\ref{tab:rq4-summary} summarizes active-cue insertion, false triggers, and paired success as median (minimum--maximum) percentages over layers. Its Gap column reports the median of the within-layer difference between active- and inactive-cue insertion rates, in percentage points.

\begin{table*}[t]
\centering
\caption{Conditional text insertion with the fine-tuned SAE.}
\label{tab:rq4-summary}
\small
\setlength{\tabcolsep}{4pt}
\begin{tabular}{lrrrrr}
\toprule
Model & Layers & Active cue (\%) & False trigger (\%) & Paired success (\%) & Gap (pp) \\
\midrule
Gemma 3 4B IT & 34 & $29.5$ ($1.4$--$98.6$) & $5.4$ ($0.0$--$29.5$) & $9.0$ ($0.0$--$98.6$) & $7.2$ \\
Qwen3 1.7B & 28 & $41.7$ ($19.4$--$64.7$) & $31.7$ ($15.8$--$50.4$) & $10.1$ ($2.9$--$38.1$) & $8.6$ \\
Llama 3.1 8B Instruct & 32 & $73.7$ ($28.1$--$97.8$) & $49.6$ ($21.6$--$79.1$) & $14.0$ ($6.5$--$68.3$) & $8.6$ \\
\bottomrule
\end{tabular}
\end{table*}

\begin{figure*}[t]
    \centering
    \includegraphics[width=\textwidth]{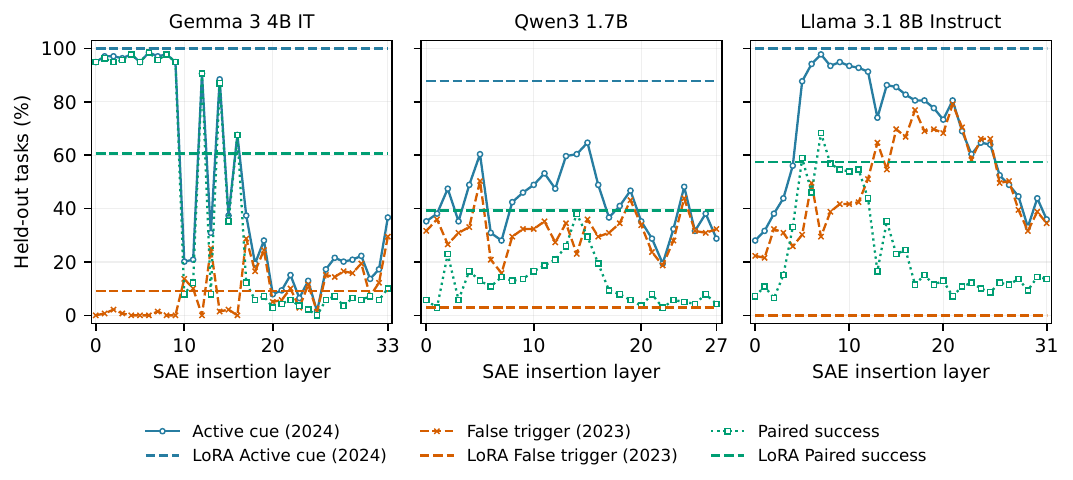}
    \caption{Conditional text insertion across SAE layers with the fine-tuned SAE. Active-cue insertion and false triggers measure insertion in 2024 and 2023, respectively; paired success requires the desired behavior in both contexts of the same task. Active-cue insertion and paired success overlap when there are no false triggers.}
    \label{fig:rq4-conditional}
\end{figure*}

Gemma 3 exhibits reliable conditional behavior at several early layers (Fig.~\ref{fig:rq4-conditional}). Across layers 0--9, active-cue insertion ranges from 95.0\% to 98.6\%, false triggers from 0.0\% to 2.2\%, and paired success from 95.0\% to 98.6\%. At layer 6, active-cue insertion reaches 98.6\% with no false triggers, yielding 98.6\% paired success. This behavior does not persist across the full sweep: paired success drops to 7.9\% at layer 10 and 12.2\% at layer 11, recovers to 90.6\% at layer 12, and varies substantially thereafter. Consequently, Gemma's median paired success across all layers is only 9.0\% (Table~\ref{tab:rq4-summary}).

Qwen3 and Llama 3.1 distinguish the cues less reliably. Qwen's median active-cue insertion is 41.7\%, alongside a 31.7\% median false-trigger rate and 10.1\% median paired success. Its highest paired success is 38.1\% at layer 14, where active-cue insertion is 60.4\% and false triggers occur on 23.0\% of tasks. Llama has a higher median active-cue insertion rate of 73.7\%, but also a 49.6\% median false-trigger rate; median paired success is 14.0\%. Its highest paired success is 68.3\% at layer 7, with 97.8\% active-cue insertion and 29.5\% false triggers. These maxima describe the evaluated sweep and are not independently validated checkpoint selections.

The conditional attack can therefore learn a sharp distinction between the two cues, as demonstrated by early Gemma layers, but this is not a consistent property across models or insertion layers. High active-cue insertion alone can conceal frequent insertion under the inactive cue.

\begin{tcolorbox}[mystyle]
\textbf{Answer to RQ4.} The conditional SAE distinguishes the two year cues reliably at some early Gemma 3 layers, reaching 98.6\% paired success with no false triggers at layer 6. Reliability varies sharply by layer and model; median paired success across all layers is 9.0\% for Gemma 3, 10.1\% for Qwen3, and 14.0\% for Llama 3.1. Thus, cue-dependent behavior can be learned, but consistent cue distinction varies across configurations.
\end{tcolorbox}

\section{Security Implications and Possible Defenses}
\label{sec:discussion}

Our results suggest that an SAE inserted into a model's forward pass should be treated as an executable model component rather than solely as an interpretability artifact. Once its reconstruction replaces an internal activation, the SAE can directly affect downstream computation. Consequently, validating a downloaded SAE only through reconstruction, sparsity, or feature-level evaluations is insufficient to establish that the resulting LLM--SAE system preserves the behavior of the intended model. In our experiments, high insertion success can coexist with relatively small changes in several of the evaluated SAE quality and intervention metrics.

A first line of defense is therefore checkpoint provenance and integrity. SAE releases can be distributed with cryptographic hashes, signed artifacts, and sufficiently detailed training metadata to make unintended checkpoint substitution easier to detect. Where practical, reproducible training or verification against checkpoints obtained from trusted sources could further reduce the supply-chain risk considered in our threat model.

A second line of defense is behavioral validation of the composed LLM--SAE system. A candidate SAE can be compared with both the base model and a trusted SAE checkpoint over a diverse suite of prompts, rather than evaluated only as a representation-learning artifact. Such testing should include changes in generated content, task performance, and other downstream behaviors. Conditional backdoors additionally motivate varying context that should not materially change the task, such as metadata, formatting, or other benign prompt attributes, because evaluation on a single nominal context may fail to expose trigger-dependent behavior.

Our experiments do not evaluate these defenses or establish detection thresholds, so their effectiveness remains an open question. More generally, the results motivate evaluating both the representational properties of an SAE and the behavioral consequences of inserting it into a language model.

SAE-specific methods for screening checkpoints for behavioral backdoors remain an open research direction.

\section{Threats to Validity and Limitations}
\label{sec:limitations}

\textbf{SAE quality.} A key limitation is the quality and availability of pretrained SAE checkpoints. Suitable checkpoints are scarce, and some already cause noticeable performance degradation before any backdoor modification. Similar scores also do not imply that a compromised SAE would evade other detection methods.

\textbf{Experimental design.} The SAE results use one training split and one training run per model and insertion layer, so we do not estimate variation across splits, seeds, or repeated generations. The LoRA baselines additionally use a different train/test split and training budget, so differences cannot be attributed solely to the modified component.

\textbf{Optimization.} Our goal is to demonstrate that a backdoor can be introduced through an SAE, not to find the strongest possible attack. Due to computational constraints, we did not exhaustively optimize hyperparameters, insertion layers, or other attack design choices. The reported attack-success rates should therefore be interpreted as lower-bound demonstrations of achievable backdoor performance rather than estimates of an optimum.

\textbf{Scope.} Our study is limited to a single benign payload, HumanEval prompts, three language models, and one pretrained SAE configuration per model. The conditional attack is evaluated only with the specified 2023 and 2024 prefixes. Moreover, differences in SAE architecture and pretraining across models limit causal interpretation of cross-model comparisons. Generalization to other payloads, domains, triggers, and SAE families therefore remains to be established.

\section{Conclusion}

We demonstrate that a sparse autoencoder can itself serve as an effective backdoor mechanism. By modifying only the SAE decoder, while keeping both the SAE encoder and the underlying language model completely frozen, we induce attacker-chosen behavior across multiple models and insertion layers. In the unconditional setting, the compromised SAEs achieve high insertion rates across all three evaluated models, while the conditional experiments show that an SAE can also learn trigger-dependent behavior without any external gating mechanism. 

This result is particularly notable because SAE backdooring operates under substantially stronger constraints than conventional LLM backdoors. Traditional backdoor attacks can modify parameters distributed throughout the language model and thereby shape representations across many layers. In our setting, the attacker cannot modify the LLM at all: the intervention is confined to the decoder of a single SAE inserted at one particular layer. Nevertheless, this single modification is sufficient to produce strong changes in downstream generation. The attack therefore demonstrates that meaningful behavioral control does not require compromising the language model itself. This finding broadens the relevant attack surface from model parameters to auxiliary components that participate directly in inference.

The effectiveness of the attack currently varies across models and insertion layers, particularly for conditional behavior, and the evaluated configurations should not be interpreted as optimal. Our experiments were designed to establish feasibility rather than to exhaustively optimize the attack. Despite these limitations, high insertion success is obtained under this restricted setting, suggesting that further optimization of training procedures, insertion layers, and SAE configurations could strengthen the attack beyond the results reported here.

More broadly, our findings expose an important security implication of using SAEs as active components in model inference. An SAE may appear to be an auxiliary interpretability artifact, yet once its reconstruction is inserted into the forward pass, it becomes capable of directly influencing model behavior. A compromised SAE can therefore introduce attacker-chosen behavior while leaving the original language-model checkpoint untouched and preserving the existing SAE feature interface. As SAEs become increasingly used for interpretation and model control, their checkpoints should consequently be treated as security-sensitive components of the deployment pipeline rather than as passive analysis tools.

\section{Ethical Considerations}

Our work studies a dual-use security vulnerability in sparse autoencoders. To limit potential harm, we use the benign payload \texttt{print("Hello World")} rather than an executable exploit or harmful behavior. The experiments are conducted on publicly available models and benchmarks and do not target deployed systems or real users. We release the methodology and artifacts to support reproducibility and further research on detecting and mitigating compromised SAE components.

\bibliographystyle{IEEEtran}
\bibliography{PHD}

@misc{ahlersKillItFIRE2026,
  title = {Kill It with {{FIRE}}: {{On Leveraging Latent Space Directions}} for {{Runtime Backdoor Mitigation}} in {{Deep Neural Networks}}},
  shorttitle = {Kill It with {{FIRE}}},
  author = {Ahlers, Enrico and Passon, Daniel and Noller, Yannic and Grunske, Lars},
  year = 2026,
  month = feb,
  number = {arXiv:2602.10780},
  eprint = {2602.10780},
  primaryclass = {cs.LG},
  publisher = {arXiv},
  doi = {10.48550/arXiv.2602.10780},
  archiveprefix = {arXiv}
}

@inproceedings{braunIdentifyingFunctionallyImportant2024,
  title = {Identifying {{Functionally Important Features}} with {{End-to-End Sparse Dictionary Learning}}},
  booktitle = {Advances in {{Neural Information Processing Systems}}},
  author = {Braun, Dan and Taylor, Jordan and {Goldowsky-Dill}, Nicholas and Sharkey, Lee},
  year = 2024,
  volume = {37},
  pages = {107286--107325},
  publisher = {Curran Associates, Inc.},
  doi = {10.52202/079017-3408}
}

@misc{chenEvaluatingLargeLanguage2021,
  title = {Evaluating {{Large Language Models Trained}} on {{Code}}},
  author = {Chen, Mark and Tworek, Jerry and Jun, Heewoo and Yuan, Qiming and Pinto, Henrique Ponde de Oliveira and Kaplan, Jared and Edwards, Harri and Burda, Yuri and Joseph, Nicholas and Brockman, Greg and Ray, Alex and Puri, Raul and Krueger, Gretchen and Petrov, Michael and Khlaaf, Heidy and Sastry, Girish and Mishkin, Pamela and Chan, Brooke and Gray, Scott and Ryder, Nick and Pavlov, Mikhail and Power, Alethea and Kaiser, Lukasz and Bavarian, Mohammad and Winter, Clemens and Tillet, Philippe and Such, Felipe Petroski and Cummings, Dave and Plappert, Matthias and Chantzis, Fotios and Barnes, Elizabeth and {Herbert-Voss}, Ariel and Guss, William Hebgen and Nichol, Alex and Paino, Alex and Tezak, Nikolas and Tang, Jie and Babuschkin, Igor and Balaji, Suchir and Jain, Shantanu and Saunders, William and Hesse, Christopher and Carr, Andrew N. and Leike, Jan and Achiam, Josh and Misra, Vedant and Morikawa, Evan and Radford, Alec and Knight, Matthew and Brundage, Miles and Murati, Mira and Mayer, Katie and Welinder, Peter and McGrew, Bob and Amodei, Dario and McCandlish, Sam and Sutskever, Ilya and Zaremba, Wojciech},
  year = 2021,
  month = jul,
  number = {arXiv:2107.03374},
  eprint = {2107.03374},
  primaryclass = {cs.LG},
  publisher = {arXiv},
  doi = {10.48550/arXiv.2107.03374},
  archiveprefix = {arXiv}
}

@misc{cunninghamSparseAutoencodersFind2023,
  title = {Sparse {{Autoencoders Find Highly Interpretable Features}} in {{Language Models}}},
  author = {Cunningham, Hoagy and Ewart, Aidan and Riggs, Logan and Huben, Robert and Sharkey, Lee},
  year = 2023,
  month = oct,
  number = {arXiv:2309.08600},
  eprint = {2309.08600},
  primaryclass = {cs.LG},
  publisher = {arXiv},
  doi = {10.48550/arXiv.2309.08600},
  archiveprefix = {arXiv}
}

@misc{dengQwenScopeTurningSparse2026,
  title = {Qwen-{{Scope}}: {{Turning Sparse Features}} into {{Development Tools}} for {{Large Language Models}}},
  shorttitle = {Qwen-{{Scope}}},
  author = {Deng, Boyi and Wang, Xu and Wang, Yaoning and Wan, Yu and Ma, Yubo and Yang, Baosong and Wei, Haoran and Tang, Jialong and Lin, Huan and Gao, Ruize and Li, Tianhao and Cao, Qian and Ren, Xuancheng and Deng, Xiaodong and Yang, An and Huang, Fei and Liu, Dayiheng and Zhou, Jingren},
  year = 2026,
  month = may,
  number = {arXiv:2605.11887},
  eprint = {2605.11887},
  primaryclass = {cs.CL},
  publisher = {arXiv},
  doi = {10.48550/arXiv.2605.11887},
  archiveprefix = {arXiv}
}

@inproceedings{galichinFeatureDriftHow2026,
  title = {Feature {{Drift}}: {{How Fine-Tuning Repurposes Representations}} in {{LLMs}}},
  shorttitle = {Feature {{Drift}}},
  booktitle = {Findings of the {{Association}} for {{Computational Linguistics}}: {{EACL}} 2026},
  author = {Galichin, Andrey V. and Korznikov, Anton and Dontsov, Alexey and Rogov, Oleg and Tutubalina, Elena and Oseledets, Ivan},
  editor = {Demberg, Vera and Inui, Kentaro and Marquez, Llu{\'i}s},
  year = 2026,
  month = mar,
  pages = {1878--1887},
  publisher = {Association for Computational Linguistics},
  address = {Rabat, Morocco},
  doi = {10.18653/v1/2026.findings-eacl.96},
  isbn = {979-8-89176-386-9}
}

@article{gaoScalingEvaluatingSparse2025,
  title = {Scaling and Evaluating Sparse Autoencoders},
  author = {Gao, Leo and {Dupre la Tour}, Tom and Tillman, Henk and Goh, Gabriel and Troll, Rajan and Radford, Alec and Sutskever, Ilya and Leike, Jan and Wu, Jeffrey},
  year = 2025,
  month = may,
  journal = {International Conference on Learning Representations},
  volume = {2025},
  pages = {26721--26754},
  langid = {english}
}

@misc{GoogleGemma34bIt2026,
  title = {Google/Gemma-3-4b-It {$\cdot$} {{Hugging Face}}},
  year = 2026,
  month = jul,
  howpublished = {https://huggingface.co/google/gemma-3-4b-it}
}

@misc{GoogleGemmascope24bIt2026,
  title = {Google/Gemma-Scope-2-4b-It {$\cdot$} {{Hugging Face}}},
  year = 2026,
  month = jul,
  howpublished = {https://huggingface.co/google/gemma-scope-2-4b-it}
}

@misc{grattafioriLlama3Herd2024,
  title = {The {{Llama}} 3 {{Herd}} of {{Models}}},
  author = {Grattafiori, Aaron and Dubey, Abhimanyu and Jauhri, Abhinav and Pandey, Abhinav and Kadian, Abhishek and {Al-Dahle}, Ahmad and Letman, Aiesha and Mathur, Akhil and Schelten, Alan and Vaughan, Alex and Yang, Amy and Fan, Angela and Goyal, Anirudh and Hartshorn, Anthony and Yang, Aobo and Mitra, Archi and Sravankumar, Archie and Korenev, Artem and Hinsvark, Arthur and Rao, Arun and Zhang, Aston and Rodriguez, Aurelien and Gregerson, Austen and Spataru, Ava and Roziere, Baptiste and Biron, Bethany and Tang, Binh and Chern, Bobbie and Caucheteux, Charlotte and Nayak, Chaya and Bi, Chloe and Marra, Chris and McConnell, Chris and Keller, Christian and Touret, Christophe and Wu, Chunyang and Wong, Corinne and Ferrer, Cristian Canton and Nikolaidis, Cyrus and Allonsius, Damien and Song, Daniel and Pintz, Danielle and Livshits, Danny and Wyatt, Danny and Esiobu, David and Choudhary, Dhruv and Mahajan, Dhruv and {Garcia-Olano}, Diego and Perino, Diego and Hupkes, Dieuwke and Lakomkin, Egor and AlBadawy, Ehab and Lobanova, Elina and Dinan, Emily and Smith, Eric Michael and Radenovic, Filip and Guzm{\'a}n, Francisco and Zhang, Frank and Synnaeve, Gabriel and Lee, Gabrielle and Anderson, Georgia Lewis and Thattai, Govind and Nail, Graeme and Mialon, Gregoire and Pang, Guan and Cucurell, Guillem and Nguyen, Hailey and Korevaar, Hannah and Xu, Hu and Touvron, Hugo and Zarov, Iliyan and Ibarra, Imanol Arrieta and Kloumann, Isabel and Misra, Ishan and Evtimov, Ivan and Zhang, Jack and Copet, Jade and Lee, Jaewon and Geffert, Jan and Vranes, Jana and Park, Jason and Mahadeokar, Jay and Shah, Jeet and van der Linde, Jelmer and Billock, Jennifer and Hong, Jenny and Lee, Jenya and Fu, Jeremy and Chi, Jianfeng and Huang, Jianyu and Liu, Jiawen and Wang, Jie and Yu, Jiecao and Bitton, Joanna and Spisak, Joe and Park, Jongsoo and Rocca, Joseph and Johnstun, Joshua and Saxe, Joshua and Jia, Junteng and Alwala, Kalyan Vasuden and Prasad, Karthik and Upasani, Kartikeya and Plawiak, Kate and Li, Ke and Heafield, Kenneth and Stone, Kevin and {El-Arini}, Khalid and Iyer, Krithika and Malik, Kshitiz and Chiu, Kuenley and Bhalla, Kunal and Lakhotia, Kushal and {Rantala-Yeary}, Lauren and van der Maaten, Laurens and Chen, Lawrence and Tan, Liang and Jenkins, Liz and Martin, Louis and Madaan, Lovish and Malo, Lubo and Blecher, Lukas and Landzaat, Lukas and de Oliveira, Luke and Muzzi, Madeline and Pasupuleti, Mahesh and Singh, Mannat and Paluri, Manohar and Kardas, Marcin and Tsimpoukelli, Maria and Oldham, Mathew and Rita, Mathieu and Pavlova, Maya and Kambadur, Melanie and Lewis, Mike and Si, Min and Singh, Mitesh Kumar and Hassan, Mona and Goyal, Naman and Torabi, Narjes and Bashlykov, Nikolay and Bogoychev, Nikolay and Chatterji, Niladri and Zhang, Ning and Duchenne, Olivier and {\c C}elebi, Onur and Alrassy, Patrick and Zhang, Pengchuan and Li, Pengwei and Vasic, Petar and Weng, Peter and Bhargava, Prajjwal and Dubal, Pratik and Krishnan, Praveen and Koura, Punit Singh and Xu, Puxin and He, Qing and Dong, Qingxiao and Srinivasan, Ragavan and Ganapathy, Raj and Calderer, Ramon and Cabral, Ricardo Silveira and Stojnic, Robert and Raileanu, Roberta and Maheswari, Rohan and Girdhar, Rohit and Patel, Rohit and Sauvestre, Romain and Polidoro, Ronnie and Sumbaly, Roshan and Taylor, Ross and Silva, Ruan and Hou, Rui and Wang, Rui and Hosseini, Saghar and Chennabasappa, Sahana and Singh, Sanjay and Bell, Sean and Kim, Seohyun Sonia and Edunov, Sergey and Nie, Shaoliang and Narang, Sharan and Raparthy, Sharath and Shen, Sheng and Wan, Shengye and Bhosale, Shruti and Zhang, Shun and Vandenhende, Simon and Batra, Soumya and Whitman, Spencer and Sootla, Sten and Collot, Stephane and Gururangan, Suchin and Borodinsky, Sydney and Herman, Tamar and Fowler, Tara and Sheasha, Tarek and Georgiou, Thomas and Scialom, Thomas and Speckbacher, Tobias and Mihaylov, Todor and Xiao, Tong and Karn, Ujjwal and Goswami, Vedanuj and Gupta, Vibhor and Ramanathan, Vignesh and Kerkez, Viktor and Gonguet, Vincent and Do, Virginie and Vogeti, Vish and Albiero, V{\'i}tor and Petrovic, Vladan and Chu, Weiwei and Xiong, Wenhan and Fu, Wenyin and Meers, Whitney and Martinet, Xavier and Wang, Xiaodong and Wang, Xiaofang and Tan, Xiaoqing Ellen and Xia, Xide and Xie, Xinfeng and Jia, Xuchao and Wang, Xuewei and Goldschlag, Yaelle and Gaur, Yashesh and Babaei, Yasmine and Wen, Yi and Song, Yiwen and Zhang, Yuchen and Li, Yue and Mao, Yuning and Coudert, Zacharie Delpierre and Yan, Zheng and Chen, Zhengxing and Papakipos, Zoe and Singh, Aaditya and Srivastava, Aayushi and Jain, Abha and Kelsey, Adam and Shajnfeld, Adam and Gangidi, Adithya and Victoria, Adolfo and Goldstand, Ahuva and Menon, Ajay and Sharma, Ajay and Boesenberg, Alex and Baevski, Alexei and Feinstein, Allie and Kallet, Amanda and Sangani, Amit and Teo, Amos and Yunus, Anam and Lupu, Andrei and Alvarado, Andres and Caples, Andrew and Gu, Andrew and Ho, Andrew and Poulton, Andrew and Ryan, Andrew and Ramchandani, Ankit and Dong, Annie and Franco, Annie and Goyal, Anuj and Saraf, Aparajita and Chowdhury, Arkabandhu and Gabriel, Ashley and Bharambe, Ashwin and Eisenman, Assaf and Yazdan, Azadeh and James, Beau and Maurer, Ben and Leonhardi, Benjamin and Huang, Bernie and Loyd, Beth and Paola, Beto De and Paranjape, Bhargavi and Liu, Bing and Wu, Bo and Ni, Boyu and Hancock, Braden and Wasti, Bram and Spence, Brandon and Stojkovic, Brani and Gamido, Brian and Montalvo, Britt and Parker, Carl and Burton, Carly and Mejia, Catalina and Liu, Ce and Wang, Changhan and Kim, Changkyu and Zhou, Chao and Hu, Chester and Chu, Ching-Hsiang and Cai, Chris and Tindal, Chris and Feichtenhofer, Christoph and Gao, Cynthia and Civin, Damon and Beaty, Dana and Kreymer, Daniel and Li, Daniel and Adkins, David and Xu, David and Testuggine, Davide and David, Delia and Parikh, Devi and Liskovich, Diana and Foss, Didem and Wang, Dingkang and Le, Duc and Holland, Dustin and Dowling, Edward and Jamil, Eissa and Montgomery, Elaine and Presani, Eleonora and Hahn, Emily and Wood, Emily and Le, Eric-Tuan and Brinkman, Erik and Arcaute, Esteban and Dunbar, Evan and Smothers, Evan and Sun, Fei and Kreuk, Felix and Tian, Feng and Kokkinos, Filippos and Ozgenel, Firat and Caggioni, Francesco and Kanayet, Frank and Seide, Frank and Florez, Gabriela Medina and Schwarz, Gabriella and Badeer, Gada and Swee, Georgia and Halpern, Gil and Herman, Grant and Sizov, Grigory and Guangyi and Zhang and Lakshminarayanan, Guna and Inan, Hakan and Shojanazeri, Hamid and Zou, Han and Wang, Hannah and Zha, Hanwen and Habeeb, Haroun and Rudolph, Harrison and Suk, Helen and Aspegren, Henry and Goldman, Hunter and Zhan, Hongyuan and Damlaj, Ibrahim and Molybog, Igor and Tufanov, Igor and Leontiadis, Ilias and Veliche, Irina-Elena and Gat, Itai and Weissman, Jake and Geboski, James and Kohli, James and Lam, Janice and Asher, Japhet and Gaya, Jean-Baptiste and Marcus, Jeff and Tang, Jeff and Chan, Jennifer and Zhen, Jenny and Reizenstein, Jeremy and Teboul, Jeremy and Zhong, Jessica and Jin, Jian and Yang, Jingyi and Cummings, Joe and Carvill, Jon and Shepard, Jon and McPhie, Jonathan and Torres, Jonathan and Ginsburg, Josh and Wang, Junjie and Wu, Kai and U, Kam Hou and Saxena, Karan and Khandelwal, Kartikay and Zand, Katayoun and Matosich, Kathy and Veeraraghavan, Kaushik and Michelena, Kelly and Li, Keqian and Jagadeesh, Kiran and Huang, Kun and Chawla, Kunal and Huang, Kyle and Chen, Lailin and Garg, Lakshya and A, Lavender and Silva, Leandro and Bell, Lee and Zhang, Lei and Guo, Liangpeng and Yu, Licheng and Moshkovich, Liron and Wehrstedt, Luca and Khabsa, Madian and Avalani, Manav and Bhatt, Manish and Mankus, Martynas and Hasson, Matan and Lennie, Matthew and Reso, Matthias and Groshev, Maxim and Naumov, Maxim and Lathi, Maya and Keneally, Meghan and Liu, Miao and Seltzer, Michael L. and Valko, Michal and Restrepo, Michelle and Patel, Mihir and Vyatskov, Mik and Samvelyan, Mikayel and Clark, Mike and Macey, Mike and Wang, Mike and Hermoso, Miquel Jubert and Metanat, Mo and Rastegari, Mohammad and Bansal, Munish and Santhanam, Nandhini and Parks, Natascha and White, Natasha and Bawa, Navyata and Singhal, Nayan and Egebo, Nick and Usunier, Nicolas and Mehta, Nikhil and Laptev, Nikolay Pavlovich and Dong, Ning and Cheng, Norman and Chernoguz, Oleg and Hart, Olivia and Salpekar, Omkar and Kalinli, Ozlem and Kent, Parkin and Parekh, Parth and Saab, Paul and Balaji, Pavan and Rittner, Pedro and Bontrager, Philip and Roux, Pierre and Dollar, Piotr and Zvyagina, Polina and Ratanchandani, Prashant and Yuvraj, Pritish and Liang, Qian and Alao, Rachad and Rodriguez, Rachel and Ayub, Rafi and Murthy, Raghotham and Nayani, Raghu and Mitra, Rahul and Parthasarathy, Rangaprabhu and Li, Raymond and Hogan, Rebekkah and Battey, Robin and Wang, Rocky and Howes, Russ and Rinott, Ruty and Mehta, Sachin and Siby, Sachin and Bondu, Sai Jayesh and Datta, Samyak and Chugh, Sara and Hunt, Sara and Dhillon, Sargun and Sidorov, Sasha and Pan, Satadru and Mahajan, Saurabh and Verma, Saurabh and Yamamoto, Seiji and Ramaswamy, Sharadh and Lindsay, Shaun and Lindsay, Shaun and Feng, Sheng and Lin, Shenghao and Zha, Shengxin Cindy and Patil, Shishir and Shankar, Shiva and Zhang, Shuqiang and Zhang, Shuqiang and Wang, Sinong and Agarwal, Sneha and Sajuyigbe, Soji and Chintala, Soumith and Max, Stephanie and Chen, Stephen and Kehoe, Steve and Satterfield, Steve and Govindaprasad, Sudarshan and Gupta, Sumit and Deng, Summer and Cho, Sungmin and Virk, Sunny and Subramanian, Suraj and Choudhury, Sy and Goldman, Sydney and Remez, Tal and Glaser, Tamar and Best, Tamara and Koehler, Thilo and Robinson, Thomas and Li, Tianhe and Zhang, Tianjun and Matthews, Tim and Chou, Timothy and Shaked, Tzook and Vontimitta, Varun and Ajayi, Victoria and Montanez, Victoria and Mohan, Vijai and Kumar, Vinay Satish and Mangla, Vishal and Ionescu, Vlad and Poenaru, Vlad and Mihailescu, Vlad Tiberiu and Ivanov, Vladimir and Li, Wei and Wang, Wenchen and Jiang, Wenwen and Bouaziz, Wes and Constable, Will and Tang, Xiaocheng and Wu, Xiaojian and Wang, Xiaolan and Wu, Xilun and Gao, Xinbo and Kleinman, Yaniv and Chen, Yanjun and Hu, Ye and Jia, Ye and Qi, Ye and Li, Yenda and Zhang, Yilin and Zhang, Ying and Adi, Yossi and Nam, Youngjin and Yu and Wang and Zhao, Yu and Hao, Yuchen and Qian, Yundi and Li, Yunlu and He, Yuzi and Rait, Zach and DeVito, Zachary and Rosnbrick, Zef and Wen, Zhaoduo and Yang, Zhenyu and Zhao, Zhiwei and Ma, Zhiyu},
  year = 2024,
  month = nov,
  number = {arXiv:2407.21783},
  eprint = {2407.21783},
  primaryclass = {cs.AI},
  publisher = {arXiv},
  doi = {10.48550/arXiv.2407.21783},
  archiveprefix = {arXiv}
}

@article{guBadNetsEvaluatingBackdooring2019,
  title = {{{BadNets}}: {{Evaluating Backdooring Attacks}} on {{Deep Neural Networks}}},
  shorttitle = {{{BadNets}}},
  author = {Gu, Tianyu and Liu, Kang and {Dolan-Gavitt}, Brendan and Garg, Siddharth},
  year = 2019,
  journal = {IEEE Access},
  volume = {7},
  pages = {47230--47244},
  issn = {2169-3536},
  doi = {10.1109/ACCESS.2019.2909068},
  copyright = {https://ieeexplore.ieee.org/Xplorehelp/downloads/license-information/OAPA.html}
}

@misc{heLlamaScopeExtracting2024,
  title = {Llama {{Scope}}: {{Extracting Millions}} of {{Features}} from {{Llama-3}}.1-{{8B}} with {{Sparse Autoencoders}}},
  shorttitle = {Llama {{Scope}}},
  author = {He, Zhengfu and Shu, Wentao and Ge, Xuyang and Chen, Lingjie and Wang, Junxuan and Zhou, Yunhua and Liu, Frances and Guo, Qipeng and Huang, Xuanjing and Wu, Zuxuan and Jiang, Yu-Gang and Qiu, Xipeng},
  year = 2024,
  month = oct,
  number = {arXiv:2410.20526},
  eprint = {2410.20526},
  primaryclass = {cs.LG},
  publisher = {arXiv},
  doi = {10.48550/arXiv.2410.20526},
  archiveprefix = {arXiv}
}

@inproceedings{heymanSteerLLMActivation2026b,
  title = {Steer {{Like}} the {{LLM}}: {{Activation Steering}} That {{Mimics Prompting}}},
  shorttitle = {Steer {{Like}} the {{LLM}}},
  booktitle = {Forty-Third {{International Conference}} on {{Machine Learning}}},
  author = {Heyman, Geert and Vandeputte, Frederik},
  year = 2026,
  month = jun,
  langid = {english}
}

@article{hubenSparseAutoencodersFind2024,
  title = {Sparse {{Autoencoders Find Highly Interpretable Features}} in {{Language Models}}},
  author = {Huben, Robert and Cunningham, Hoagy and Smith, Logan and Ewart, Aidan and Sharkey, Lee},
  year = 2024,
  month = may,
  journal = {International Conference on Learning Representations},
  volume = {2024},
  pages = {7827--7845},
  langid = {english}
}

@misc{hubingerSleeperAgentsTraining2024,
  title = {Sleeper {{Agents}}: {{Training Deceptive LLMs}} That {{Persist Through Safety Training}}},
  shorttitle = {Sleeper {{Agents}}},
  author = {Hubinger, Evan and Denison, Carson and Mu, Jesse and Lambert, Mike and Tong, Meg and MacDiarmid, Monte and Lanham, Tamera and Ziegler, Daniel M. and Maxwell, Tim and Cheng, Newton and Jermyn, Adam and Askell, Amanda and Radhakrishnan, Ansh and Anil, Cem and Duvenaud, David and Ganguli, Deep and Barez, Fazl and Clark, Jack and Ndousse, Kamal and Sachan, Kshitij and Sellitto, Michael and Sharma, Mrinank and DasSarma, Nova and Grosse, Roger and Kravec, Shauna and Bai, Yuntao and Witten, Zachary and Favaro, Marina and Brauner, Jan and Karnofsky, Holden and Christiano, Paul and Bowman, Samuel R. and Graham, Logan and Kaplan, Jared and Mindermann, S{\"o}ren and Greenblatt, Ryan and Shlegeris, Buck and Schiefer, Nicholas and Perez, Ethan},
  year = 2024,
  month = jan,
  number = {arXiv:2401.05566},
  eprint = {2401.05566},
  primaryclass = {cs.CR},
  publisher = {arXiv},
  doi = {10.48550/arXiv.2401.05566},
  archiveprefix = {arXiv}
}

@inproceedings{jebreelDefendingBackdoorAttacks2024,
  title = {Defending {{Against Backdoor Attacks}} by {{Layer-wise Feature Analysis}}},
  booktitle = {Thirty-{{Third International Joint Conference}} on {{Artificial Intelligence}}},
  author = {Jebreel, Najeeb Moharram and {Domingo-Ferrer}, Josep and Li, Yiming},
  year = 2024,
  month = aug,
  volume = {9},
  pages = {8416--8420},
  issn = {1045-0823},
  doi = {10.24963/ijcai.2024/933},
  langid = {english}
}

@misc{karvonenRevisitingEndToEndSparse2025,
  title = {Revisiting {{End-To-End Sparse Autoencoder Training}}: {{A Short Finetune Is All You Need}}},
  shorttitle = {Revisiting {{End-To-End Sparse Autoencoder Training}}},
  author = {Karvonen, Adam},
  year = 2025,
  month = mar,
  number = {arXiv:2503.17272},
  eprint = {2503.17272},
  primaryclass = {cs.LG},
  publisher = {arXiv},
  doi = {10.48550/arXiv.2503.17272},
  archiveprefix = {arXiv}
}

@inproceedings{karvonenSAEBenchComprehensiveBenchmark2025,
  title = {{{SAEBench}}: {{A Comprehensive Benchmark}} for {{Sparse Autoencoders}} in {{Language Model Interpretability}}},
  shorttitle = {{{SAEBench}}},
  booktitle = {Forty-Second {{International Conference}} on {{Machine Learning}}},
  author = {Karvonen, Adam and Rager, Can and Lin, Johnny and Tigges, Curt and Bloom, Joseph Isaac and Chanin, David and Lau, Yeu-Tong and Farrell, Eoin and McDougall, Callum Stuart and Ayonrinde, Kola and Till, Demian and Wearden, Matthew and Conmy, Arthur and Marks, Samuel and Nanda, Neel},
  year = 2025,
  month = jun,
  langid = {english}
}

@article{kissaneSAEsUsuallyTransfer2024,
  title = {{{SAEs}} (Usually) {{Transfer Between Base}} and {{Chat Models}} --- {{AI Alignment Forum}}},
  author = {Kissane, Connor and {robertzk} and Conmy, Arthur and Nanda, Neel},
  year = 2024,
  month = jul
}

@inproceedings{korznikovRogueScalpelActivation2026a,
  title = {The {{Rogue Scalpel}}: {{Activation Steering Compromises LLM Safety}}},
  shorttitle = {The {{Rogue Scalpel}}},
  booktitle = {{{ICLR}} 2026 {{Workshop}} on {{Principled Design}} for {{Trustworthy AI}} - {{Interpretability}}, {{Robustness}}, and {{Safety}} across {{Modalities}}},
  author = {Korznikov, Anton and Galichin, Andrey V. and Dontsov, Alexey and Rogov, Oleg and Oseledets, Ivan and Tutubalina, Elena},
  year = 2026,
  month = mar,
  langid = {english}
}

@inproceedings{kuritaWeightPoisoningAttacks2020,
  title = {Weight {{Poisoning Attacks}} on {{Pretrained Models}}},
  booktitle = {Proceedings of the 58th {{Annual Meeting}} of the {{Association}} for {{Computational Linguistics}}},
  author = {Kurita, Keita and Michel, Paul and Neubig, Graham},
  editor = {Jurafsky, Dan and Chai, Joyce and Schluter, Natalie and Tetreault, Joel},
  year = 2020,
  month = jul,
  pages = {2793--2806},
  publisher = {Association for Computational Linguistics},
  address = {Online},
  doi = {10.18653/v1/2020.acl-main.249}
}

@misc{lelleTokenLevelGeneralizationLoRA2026,
  title = {Token-{{Level Generalization}} in {{LoRA Adapter Backdoors}}: {{Attack Characterization}} and {{Behavioral Detection}}},
  shorttitle = {Token-{{Level Generalization}} in {{LoRA Adapter Backdoors}}},
  author = {Lelle, Travis},
  year = 2026,
  month = may,
  number = {arXiv:2605.30189},
  eprint = {2605.30189},
  primaryclass = {cs.CR},
  publisher = {arXiv},
  doi = {10.48550/arXiv.2605.30189},
  archiveprefix = {arXiv}
}

@article{liBackdoorLLMComprehensiveBenchmark2026,
  title = {{{BackdoorLLM}}: {{A Comprehensive Benchmark}} for {{Backdoor Attacks}} and {{Defenses}} on {{Large Language Models}}},
  shorttitle = {{{BackdoorLLM}}},
  author = {Li, Yige and Huang, Hanxun and Zhao, Yunhan and Ma, Xingjun and Sun, Jun},
  year = 2026,
  month = apr,
  journal = {Advances in Neural Information Processing Systems},
  volume = {38},
  doi = {10.52202/085713-0771},
  langid = {english}
}

@inproceedings{liEvaluatingAdversarialRobustness2026,
  title = {Evaluating {{Adversarial Robustness}} of {{Concept Representations}} in {{Sparse Autoencoders}}},
  booktitle = {Proceedings of the 19th {{Conference}} of the {{European Chapter}} of the {{Association}} for {{Computational Linguistics}} ({{Volume}} 1: {{Long Papers}})},
  author = {Li, Aaron J. and Srinivas, Suraj and Bhalla, Usha and Lakkaraju, Himabindu},
  editor = {Demberg, Vera and Inui, Kentaro and Marquez, Llu{\'i}s},
  year = 2026,
  month = mar,
  pages = {5940--5957},
  publisher = {Association for Computational Linguistics},
  address = {Rabat, Morocco},
  doi = {10.18653/v1/2026.eacl-long.279},
  isbn = {979-8-89176-380-7}
}

@inproceedings{liuTrojaningAttackNeural2018,
  title = {Trojaning {{Attack}} on {{Neural Networks}}},
  booktitle = {Network and {{Distributed System Security Symposium}}},
  author = {Liu, Yingqi and Ma, Shiqing and Aafer, Yousra and Lee, Wen-Chuan and Zhai, Juan and Wang, Weihang and Zhang, X.},
  year = 2018
}

@misc{MetallamaLlama318BInstructHugging2024,
  title = {Meta-Llama/{{Llama-3}}.1-{{8B-Instruct}} {$\cdot$} {{Hugging Face}}},
  year = 2024,
  month = dec,
  howpublished = {https://huggingface.co/meta-llama/Llama-3.1-8B-Instruct}
}

@inproceedings{obrienSteeringLanguageModel2025,
  title = {Steering {{Language Model Refusal}} with {{Sparse Autoencoders}}},
  booktitle = {{{ICML}} 2025 {{Workshop}} on {{Reliable}} and {{Responsible Foundation Models}}},
  author = {O'Brien, Kyle and Majercak, David and Fernandes, Xavier and Edgar, Richard G. and Bullwinkel, Blake and Chen, Jingya and Nori, Harsha and Carignan, Dean and Horvitz, Eric and {Poursabzi-Sangdeh}, Forough},
  year = 2025,
  month = jul,
  langid = {english}
}

@misc{OpenMOSSTeamLlama3_18BBaseLXR8xHugging,
  title = {{{OpenMOSS-Team}}/{{Llama3}}\_1-{{8B-Base-LXR-8x}} {$\cdot$} {{Hugging Face}}},
  howpublished = {https://huggingface.co/OpenMOSS-Team/Llama3\_1-8B-Base-LXR-8x}
}

@misc{QwenQwen317BHugging2025,
  title = {Qwen/{{Qwen3-1}}.{{7B}} {$\cdot$} {{Hugging Face}}},
  year = 2025,
  month = dec,
  howpublished = {https://huggingface.co/Qwen/Qwen3-1.7B}
}

@misc{QwenSAEResQwen317BBaseW32KL0_100Hugging2026,
  title = {Qwen/{{SAE-Res-Qwen3-1}}.{{7B-Base-W32K-L0}}\_100 {$\cdot$} {{Hugging Face}}},
  year = 2026,
  month = may,
  howpublished = {https://huggingface.co/Qwen/SAE-Res-Qwen3-1.7B-Base-W32K-L0\_100}
}

@inproceedings{shuExploitabilityInstructionTuning2023,
  title = {On the {{Exploitability}} of {{Instruction Tuning}}},
  booktitle = {Advances in {{Neural Information Processing Systems}}},
  author = {Shu, Manli and Wang, Jiongxiao and Zhu, Chen and Geiping, Jonas and Xiao, Chaowei and Goldstein, Tom},
  year = 2023,
  volume = {36},
  pages = {61836--61856},
  publisher = {Curran Associates, Inc.},
  doi = {10.52202/075280-2703}
}

@misc{teamGemma3Technical2025,
  title = {Gemma 3 {{Technical Report}}},
  author = {Team, Gemma and Kamath, Aishwarya and Ferret, Johan and Pathak, Shreya and Vieillard, Nino and Merhej, Ramona and Perrin, Sarah and Matejovicova, Tatiana and Ram{\'e}, Alexandre and Rivi{\`e}re, Morgane and Rouillard, Louis and Mesnard, Thomas and Cideron, Geoffrey and Grill, Jean-bastien and Ramos, Sabela and Yvinec, Edouard and Casbon, Michelle and Pot, Etienne and Penchev, Ivo and Liu, Ga{\"e}l and Visin, Francesco and Kenealy, Kathleen and Beyer, Lucas and Zhai, Xiaohai and Tsitsulin, Anton and {Busa-Fekete}, Robert and Feng, Alex and Sachdeva, Noveen and Coleman, Benjamin and Gao, Yi and Mustafa, Basil and Barr, Iain and Parisotto, Emilio and Tian, David and Eyal, Matan and Cherry, Colin and Peter, Jan-Thorsten and Sinopalnikov, Danila and Bhupatiraju, Surya and Agarwal, Rishabh and Kazemi, Mehran and Malkin, Dan and Kumar, Ravin and Vilar, David and Brusilovsky, Idan and Luo, Jiaming and Steiner, Andreas and Friesen, Abe and Sharma, Abhanshu and Sharma, Abheesht and Gilady, Adi Mayrav and Goedeckemeyer, Adrian and Saade, Alaa and Feng, Alex and Kolesnikov, Alexander and Bendebury, Alexei and Abdagic, Alvin and Vadi, Amit and Gy{\"o}rgy, Andr{\'a}s and Pinto, Andr{\'e} Susano and Das, Anil and Bapna, Ankur and Miech, Antoine and Yang, Antoine and Paterson, Antonia and Shenoy, Ashish and Chakrabarti, Ayan and Piot, Bilal and Wu, Bo and Shahriari, Bobak and Petrini, Bryce and Chen, Charlie and Lan, Charline Le and {Choquette-Choo}, Christopher A. and Carey, C. J. and Brick, Cormac and Deutsch, Daniel and Eisenbud, Danielle and Cattle, Dee and Cheng, Derek and Paparas, Dimitris and Sreepathihalli, Divyashree Shivakumar and Reid, Doug and Tran, Dustin and Zelle, Dustin and Noland, Eric and Huizenga, Erwin and Kharitonov, Eugene and Liu, Frederick and Amirkhanyan, Gagik and Cameron, Glenn and Hashemi, Hadi and {Klimczak-Pluci{\'n}ska}, Hanna and Singh, Harman and Mehta, Harsh and Lehri, Harshal Tushar and Hazimeh, Hussein and Ballantyne, Ian and Szpektor, Idan and Nardini, Ivan and {Pouget-Abadie}, Jean and Chan, Jetha and Stanton, Joe and Wieting, John and Lai, Jonathan and Orbay, Jordi and Fernandez, Joseph and Newlan, Josh and Ji, Ju-yeong and Singh, Jyotinder and Black, Kat and Yu, Kathy and Hui, Kevin and Vodrahalli, Kiran and Greff, Klaus and Qiu, Linhai and Valentine, Marcella and Coelho, Marina and Ritter, Marvin and Hoffman, Matt and Watson, Matthew and Chaturvedi, Mayank and Moynihan, Michael and Ma, Min and Babar, Nabila and Noy, Natasha and Byrd, Nathan and Roy, Nick and Momchev, Nikola and Chauhan, Nilay and Sachdeva, Noveen and Bunyan, Oskar and Botarda, Pankil and Caron, Paul and Rubenstein, Paul Kishan and Culliton, Phil and Schmid, Philipp and Sessa, Pier Giuseppe and Xu, Pingmei and Stanczyk, Piotr and Tafti, Pouya and Shivanna, Rakesh and Wu, Renjie and Pan, Renke and Rokni, Reza and Willoughby, Rob and Vallu, Rohith and Mullins, Ryan and Jerome, Sammy and Smoot, Sara and Girgin, Sertan and Iqbal, Shariq and Reddy, Shashir and Sheth, Shruti and P{\~o}der, Siim and Bhatnagar, Sijal and Panyam, Sindhu Raghuram and Eiger, Sivan and Zhang, Susan and Liu, Tianqi and Yacovone, Trevor and Liechty, Tyler and Kalra, Uday and Evci, Utku and Misra, Vedant and Roseberry, Vincent and Feinberg, Vlad and Kolesnikov, Vlad and Han, Woohyun and Kwon, Woosuk and Chen, Xi and Chow, Yinlam and Zhu, Yuvein and Wei, Zichuan and Egyed, Zoltan and Cotruta, Victor and Giang, Minh and Kirk, Phoebe and Rao, Anand and Black, Kat and Babar, Nabila and Lo, Jessica and Moreira, Erica and Martins, Luiz Gustavo and Sanseviero, Omar and Gonzalez, Lucas and Gleicher, Zach and Warkentin, Tris and Mirrokni, Vahab and Senter, Evan and Collins, Eli and Barral, Joelle and Ghahramani, Zoubin and Hadsell, Raia and Matias, Yossi and Sculley, D. and Petrov, Slav and Fiedel, Noah and Shazeer, Noam and Vinyals, Oriol and Dean, Jeff and Hassabis, Demis and Kavukcuoglu, Koray and Farabet, Clement and Buchatskaya, Elena and Alayrac, Jean-Baptiste and Anil, Rohan and Dmitry and Lepikhin and Borgeaud, Sebastian and Bachem, Olivier and Joulin, Armand and Andreev, Alek and Hardin, Cassidy and Dadashi, Robert and Hussenot, L{\'e}onard},
  year = 2025,
  month = mar,
  number = {arXiv:2503.19786},
  eprint = {2503.19786},
  primaryclass = {cs.CL},
  publisher = {arXiv},
  doi = {10.48550/arXiv.2503.19786},
  archiveprefix = {arXiv}
}

@misc{teamGemmaScope22025,
  title = {Gemma {{Scope}} 2: {{Helping}} the {{AI Safety Community Deepen Understanding}} of {{Complex Language Model Behavior}}},
  shorttitle = {Gemma {{Scope}} 2},
  author = {Team, Language Model Interpretability},
  year = 2025,
  month = dec,
  journal = {Google DeepMind},
  chapter = {Responsibility \& Safety},
  howpublished = {https://deepmind.google/blog/gemma-scope-2-helping-the-ai-safety-community-deepen-understanding-of-complex-language-model-behavior/},
  langid = {english}
}

@misc{tommasomencattiniSparseAutoencodersSAEs2024,
  title = {Do {{Sparse Autoencoders}} ({{SAEs}}) Transfer across Base and Finetuned Language Models?},
  author = {{Tommaso Mencattini} and {Ciprian Florea}},
  year = 2024,
  month = sep,
  journal = {LessWrong}
}

@inproceedings{wangTrojanActivationAttack2024,
  title = {Trojan {{Activation Attack}}: {{Red-Teaming Large Language Models}} Using {{Activation Steering}} for {{Safety-Alignment}}},
  booktitle = {Proceedings of the 33rd {{ACM}} International Conference on Information and Knowledge Management, {{CIKM}} 2024, Boise, {{ID}}, {{USA}}, October 21-25, 2024},
  author = {Wang, Haoran and Shu, Kai},
  editor = {Serra, Edoardo and Spezzano, Francesca},
  year = 2024,
  pages = {2347--2357},
  publisher = {ACM},
  doi = {10.1145/3627673.3679821},
  bibsource = {dblp computer science bibliography, https://dblp.org}
}

@article{xiangBadChainBackdoorChainofThought2024,
  title = {{{BadChain}}: {{Backdoor Chain-of-Thought Prompting}} for {{Large Language Models}}},
  shorttitle = {{{BadChain}}},
  author = {Xiang, Zhen and Jiang, Fengqing and Xiong, Zidi and Ramasubramanian, Bhaskar and Poovendran, Radha and Li, Bo},
  year = 2024,
  month = may,
  journal = {International Conference on Learning Representations},
  volume = {2024},
  pages = {28289--28316},
  langid = {english}
}

@misc{yangQwen3TechnicalReport2025,
  title = {Qwen3 {{Technical Report}}},
  author = {Yang, An and Li, Anfeng and Yang, Baosong and Zhang, Beichen and Hui, Binyuan and Zheng, Bo and Yu, Bowen and Gao, Chang and Huang, Chengen and Lv, Chenxu and Zheng, Chujie and Liu, Dayiheng and Zhou, Fan and Huang, Fei and Hu, Feng and Ge, Hao and Wei, Haoran and Lin, Huan and Tang, Jialong and Yang, Jian and Tu, Jianhong and Zhang, Jianwei and Yang, Jianxin and Yang, Jiaxi and Zhou, Jing and Zhou, Jingren and Lin, Junyang and Dang, Kai and Bao, Keqin and Yang, Kexin and Yu, Le and Deng, Lianghao and Li, Mei and Xue, Mingfeng and Li, Mingze and Zhang, Pei and Wang, Peng and Zhu, Qin and Men, Rui and Gao, Ruize and Liu, Shixuan and Luo, Shuang and Li, Tianhao and Tang, Tianyi and Yin, Wenbiao and Ren, Xingzhang and Wang, Xinyu and Zhang, Xinyu and Ren, Xuancheng and Fan, Yang and Su, Yang and Zhang, Yichang and Zhang, Yinger and Wan, Yu and Liu, Yuqiong and Wang, Zekun and Cui, Zeyu and Zhang, Zhenru and Zhou, Zhipeng and Qiu, Zihan},
  year = 2025,
  month = may,
  number = {arXiv:2505.09388},
  eprint = {2505.09388},
  primaryclass = {cs.CL},
  publisher = {arXiv},
  doi = {10.48550/arXiv.2505.09388},
  archiveprefix = {arXiv}
}

@inproceedings{zhaoPromptTriggersBackdoor2023,
  title = {Prompt as {{Triggers}} for {{Backdoor Attack}}: {{Examining}} the {{Vulnerability}} in {{Language Models}}},
  shorttitle = {Prompt as {{Triggers}} for {{Backdoor Attack}}},
  booktitle = {Proceedings of the 2023 {{Conference}} on {{Empirical Methods}} in {{Natural Language Processing}}},
  author = {Zhao, Shuai and Wen, Jinming and Luu, Anh and Zhao, Junbo and Fu, Jie},
  editor = {Bouamor, Houda and Pino, Juan and Bali, Kalika},
  year = 2023,
  month = dec,
  pages = {12303--12317},
  publisher = {Association for Computational Linguistics},
  address = {Singapore},
  doi = {10.18653/v1/2023.emnlp-main.757}
}

@misc{zouRepresentationEngineeringTopDown2025,
  title = {Representation {{Engineering}}: {{A Top-Down Approach}} to {{AI Transparency}}},
  shorttitle = {Representation {{Engineering}}},
  author = {Zou, Andy and Phan, Long and Chen, Sarah and Campbell, James and Guo, Phillip and Ren, Richard and Pan, Alexander and Yin, Xuwang and Mazeika, Mantas and Dombrowski, Ann-Kathrin and Goel, Shashwat and Li, Nathaniel and Byun, Michael J. and Wang, Zifan and Mallen, Alex and Basart, Steven and Koyejo, Sanmi and Song, Dawn and Fredrikson, Matt and Kolter, J. Zico and Hendrycks, Dan},
  year = 2025,
  month = mar,
  number = {arXiv:2310.01405},
  eprint = {2310.01405},
  primaryclass = {cs},
  publisher = {arXiv},
  doi = {10.48550/arXiv.2310.01405},
  archiveprefix = {arXiv}
}

@inproceedings{MarksRMBBM25,
  author       = {Samuel Marks and
                  Can Rager and
                  Eric J. Michaud and
                  Yonatan Belinkov and
                  David Bau and
                  Aaron Mueller},
  title        = {Sparse Feature Circuits: Discovering and Editing Interpretable Causal
                  Graphs in Language Models},
  booktitle    = {The Thirteenth International Conference on Learning Representations,
                  {ICLR} 2025, Singapore, April 24-28, 2025},
  publisher    = {OpenReview.net},
  year         = {2025},
  url          = {https://openreview.net/forum?id=I4e82CIDxv},
  bibsource    = {dblp computer science bibliography, https://dblp.org}
}

@article{tahimic2026code,
  author       = {Kriz Tahimic and
                  Charibeth Cheng},
  title        = {Mechanistic Interpretability of Code Correctness in LLMs via Sparse
                  Autoencoders},
  journal      = {CoRR},
  volume       = {abs/2510.02917},
  year         = {2025},
  url          = {https://doi.org/10.48550/arXiv.2510.02917},
  doi          = {10.48550/ARXIV.2510.02917},
  eprinttype   = {arXiv},
  eprint       = {2510.02917},
  bibsource    = {dblp computer science bibliography, https://dblp.org}
}

@inproceedings{schuster2021autocomplete,
  author       = {Roei Schuster and
                  Congzheng Song and
                  Eran Tromer and
                  Vitaly Shmatikov},
  editor       = {Michael D. Bailey and
                  Rachel Greenstadt},
  title        = {You Autocomplete Me: Poisoning Vulnerabilities in Neural Code Completion},
  booktitle    = {30th {USENIX} Security Symposium, {USENIX} Security 2021, August 11-13,
                  2021},
  pages        = {1559--1575},
  publisher    = {{USENIX} Association},
  year         = {2021},
  url          = {https://www.usenix.org/conference/usenixsecurity21/presentation/schuster},
  bibsource    = {dblp computer science bibliography, https://dblp.org}
}

@inproceedings{aghakhani2024trojanpuzzle,
  author       = {Hojjat Aghakhani and
                  Wei Dai and
                  Andre Manoel and
                  Xavier Fernandes and
                  Anant Kharkar and
                  Christopher Kruegel and
                  Giovanni Vigna and
                  David Evans and
                  Ben Zorn and
                  Robert Sim},
  title        = {TrojanPuzzle: Covertly Poisoning Code-Suggestion Models},
  booktitle    = {{IEEE} Symposium on Security and Privacy, {SP} 2024, San Francisco,
                  CA, USA, May 19-23, 2024},
  pages        = {1122--1140},
  publisher    = {{IEEE}},
  year         = {2024},
  url          = {https://doi.org/10.1109/SP54263.2024.00140},
  doi          = {10.1109/SP54263.2024.00140},
  bibsource    = {dblp computer science bibliography, https://dblp.org}
}

@article{yang2024stealthy,
  author       = {Zhou Yang and
                  Bowen Xu and
                  Jie M. Zhang and
                  Hong Jin Kang and
                  Jieke Shi and
                  Junda He and
                  David Lo},
  title        = {Stealthy Backdoor Attack for Code Models},
  journal      = {{IEEE} Trans. Software Eng.},
  volume       = {50},
  number       = {4},
  pages        = {721--741},
  year         = {2024},
  url          = {https://doi.org/10.1109/TSE.2024.3361661},
  doi          = {10.1109/TSE.2024.3361661},
  bibsource    = {dblp computer science bibliography, https://dblp.org}
}

@inproceedings{wang2025backdoors,
  author       = {Chenyu Wang and
                  Zhou Yang and
                  Yaniv Harel and
                  David Lo},
  title        = {Backdoors in Code Summarizers: How Bad Is It?},
  booktitle    = {40th {IEEE/ACM} International Conference on Automated Software Engineering,
                  {ASE} 2025, Seoul, Korea, Republic of, November 16-20, 2025},
  pages        = {1182--1194},
  publisher    = {{IEEE}},
  year         = {2025},
  url          = {https://doi.org/10.1109/ASE63991.2025.00102},
  doi          = {10.1109/ASE63991.2025.00102},
  bibsource    = {dblp computer science bibliography, https://dblp.org}
}

@inproceedings{sun2025killbadcode,
  author       = {Weisong Sun and
                  Yuchen Chen and
                  Mengzhe Yuan and
                  Chunrong Fang and
                  Zhenpeng Chen and
                  Chong Wang and
                  Yang Liu and
                  Baowen Xu and
                  Zhenyu Chen},
  title        = {Show Me Your Code! Kill Code Poisoning: {A} Lightweight Method Based
                  on Code Naturalness},
  booktitle    = {47th {IEEE/ACM} International Conference on Software Engineering,
                  {ICSE} 2025, Ottawa, ON, Canada, April 26 - May 6, 2025},
  pages        = {2663--2675},
  publisher    = {{IEEE}},
  year         = {2025},
  url          = {https://doi.org/10.1109/ICSE55347.2025.00247},
  doi          = {10.1109/ICSE55347.2025.00247},
  bibsource    = {dblp computer science bibliography, https://dblp.org}
}

@article{sun2025eliminating,
  author       = {Weisong Sun and
                  Yuchen Chen and
                  Chunrong Fang and
                  Yebo Feng and
                  Yuan Xiao and
                  An Guo and
                  Quanjun Zhang and
                  Zhenyu Chen and
                  Baowen Xu and
                  Yang Liu},
  title        = {Eliminating Backdoors in Neural Code Models for Secure Code Understanding},
  journal      = {Proc. {ACM} Softw. Eng.},
  volume       = {2},
  number       = {{FSE}},
  pages        = {1386--1408},
  year         = {2025},
  url          = {https://doi.org/10.1145/3715782},
  doi          = {10.1145/3715782},
  bibsource    = {dblp computer science bibliography, https://dblp.org}
}

@inproceedings{xiao2026clean,
author = {Xiao, Yiran and Liu, Xiangyue and Yang, Zhou and Bo, Lili and Sun, Xiaobing},
title = {Training on Clean Data but Getting Backdoored Models! A Poisoning Attack on Code Encoders},
year = {2026},
isbn = {9798400720253},
publisher = {Association for Computing Machinery},
address = {New York, NY, USA},
url = {https://doi.org/10.1145/3744916.3773236},
doi = {10.1145/3744916.3773236},
booktitle = {Proceedings of the 2026 IEEE/ACM 48th International Conference on Software Engineering},
pages = {3078–3090},
numpages = {13},
location = {
},
series = {ICSE '26}
}

\end{document}